\documentclass[preprint,11pt]{JHEP3}
\usepackage{epsfig,multicol}

\newcommand\fverb{\setbox\pippobox=\hbox\bgroup\verb}
\newcommand\fverbdo{\egroup\medskip\noindent%
            \fbox{\unhbox\pippobox}\ }
\newcommand\fverbit{\egroup\item[\fbox{\unhbox\pippobox}]}
\newbox\pippobox
\newcommand{\cc}{\cite}

\newcommand{\be}{\begin{equation}}
\newcommand{\ee}{\end{equation}}

\newcommand{\vecc}[1]{\mbox{\boldmath $#1$}}

\def\ve{\varepsilon}
\def\w{\omega}
\def\pd{\partial}
\def\f{\phi}
\def\F{\Phi}\def\pd{\partial}
\def\f{\phi}
\def\F{\Phi}
\def\L{\Lambda}

\def\M{\bar\m}
\def\ex{\hbox{e}}

\def\F{\Phi}
\def\<{\langle}
\def\>{\rangle}

\def\ch{\cosh}

\def\a{\alpha}
\def\b{\beta}
\def\g{\gamma}  \def\G{\Gamma}
\def\d{\delta}  \def\D{\Delta}
\def\l{\lambda}
\def\s{\sigma}
\def\r{\rho}  
\def\x{\xi}
\def\c{\chi}
\def\m{\mu}
\def\n{\nu}
\def\t{\tau}

\def\vf{\varphi}
\def\({\left(}
\def\[{\left[}
\def\){\right)}
\def\]{\right]}
\def\coth{\hbox{coth}}
\def\cot{\hbox{cot}}
\def\cos{\hbox{cos}}
\def\sin{\hbox{sin}}

\def\Tr{\hbox{Tr}}
\def\pa{{\cal P}}
\def\w1{W^{(1)}}
\def\v1{V^{(1)}}
\def\prop{D_{\m\n}}

\def\dI{\int \! dn(\r)}

\def\dk{{d^n k \over (2\pi)^n}}

\title{Instanton effects in quark form factor and quark-quark scattering at high energy}
\author{A.E. Dorokhov, I.O. Cherednikov \\
    Joint Institute for Nuclear Research \\
    RU-141980 BLTP JINR, Dubna, Russia\\
    E-mail: \email{dorokhov@thsun1.jinr.ru},
    \email{igor.cherednikov@jinr.ru} }

\preprint{2004/04/05}

\abstract{The nonperturbative effects in the high-energy processes
involving strongly interacting particles are studied within the
instanton liquid model of the QCD vacuum (ILM) by using the Wilson
integral framework. The detailed analysis of nonperturbative
contributions to the {\it electromagnetic quark form factor} is
presented considering the structure of the instanton induced
effects in the evolution equation describing the high energy
behaviour of the form factor. It is shown that the instantons
yield in high energy limit the logarithmic corrections to the
amplitudes which are exponentiated in small instanton density
parameter. By using the Gaussian interpolation of the constrained
instanton solution, we show that the all-order multi-instanton
contribution is well approximated by the weak field limit result.
The role of the instantons in {\it high energy diffractive
quark-quark scattering}, in particular, in formation of the soft
Pomeron, is also considered. We show that within the ILM the
$C-$odd diffractive amplitude is suppressed as $1/s$ compared to
the $C-$even one. The further applications of the developed
approach in studying the nonperturbative effects in high energy
hadronic processes are briefly discussed.}


\begin{document}

\section{Introduction}

The powerful methods of the perturbative Quantum ChromoDynamics
(pQCD) have been developed in order to describe the processes
involving strongly interacting particles at high energies (for a
review and comprehensive description of the methods, see, {\it
e.g.}, \cc{pQCD}). The total cross section of the $e^+e^-$
annihilation and the logarithmic violation of scaling in deep
inelastic scattering became the classical tests of pQCD already in
the lowest orders of expansion in the strong coupling constant
\cc{HARD}, and nowadays there are no doubts that the QCD
Lagrangian provides a proper basis for a quantum field theory of
strong interactions.

At the same time, the present status of pQCD does not allow one to
consider it as the only tool for investigation of the hadronic
properties even at highest energies accessible at modern machines.
The perturbative methods should be supplied by certain information
that can not be obtained directly
from pQCD calculations. For instance, a nontrivial situation
arises if, in order to make predictions of pQCD reliable, it is
necessary to resum the soft part of the quark-gluon interaction to
all orders. Moreover, in several situations the applicability of
pQCD can be definitely justified only at asymptotically high
energies, while in experimentally accessible region the
nonperturbative effects are rather important and even dominant.

Meanwhile, the intermediate energy regime---where $Q^2$ is larger
than the typical hadronic scale determined by $\L^2_{QCD}$ and
lower than the characteristic scale of the chiral symmetry
breaking: $\L^2_{QCD} \sim \L^2_{conf} < Q^2 < \L^2_{\c SB}$, is
more convenient for detection of the nonperturbative
phenomena. The complicated interplay of nonperturbative effects can lead in this
regime to formation of the constituent quark which is, in a sense,
an intermediate object between color-neutral hadron and pointlike
structureless partons, associated with the fundamental QCD
particles---quarks and gluons. Indeed, the form factors of constituent
quarks were recently extracted \cc{SCAL} from the $JLab$ experiment data
\cc{JLAB} on the inelastic Nachtmann moments \cc{NAC} of the
unpolarized proton structure function $F^p_2 (x, Q^2)$. And it was found that
the size of a constituent quark determined from
the data is about $0.2 - 0.3 \ fm$ that corresponds to the mean
instanton size in the instanton liquid model (ILM) \cc{ILM, DP}
which is also about $0.3 \ fm $.
In addition, the experimental and
phenomenological investigations of the electromagnetic quark form
factors at low and moderate energies can shed light on the
problem of scaling violation in deep inelastic scattering processes.

The requirement of certain nonperturbative supplement for explicit perturbative
calculations even in high energy domain as well the study of the hadron
processes at low and moderate energies appear to be a natural
ground for development of nonperturbative methods. Coming down in
the energy scale more and more powers of the strong coupling constant has
to be taken into account. Moreover, in the intermediate energy
region the power corrections come into play with coefficients that are very
sensitive to the intrinsic hadron structure. Typically, the
coefficients of the expansion in powers of the coupling constant
and the inverse momentum transfer squared are the hadronic
matrix elements of quark-gluon operators normalized at low energy scale and have to be
found by nonperturbative methods. The dependence of these matrix
elements on the energy scale is governed by the evolution
equations that are determined within pQCD for different hard
processes. These equations start to be applicable at momentum
transfer squared of order $1 \ GeV^{2}$ or higher where the strong
coupling constant becomes small. So, it is necessary to find the
initial data for the evolution equation that is essentially a
nonperturbative problem. The important nonperturbative
effects in this energy region are treated by different approaches:
QCD sum rules, lattice QCD, quark models, {\it etc}.
Therefore, the study of the role of the nonperturbative input in
investigation of the processes with strongly interacting particles
is not only an interesting theoretical problem, but is an
important task for phenomenology of hadronic physics.

It is naturally to relate the nonperturbative effects to the
nontrivial structure of the QCD vacuum. In the last decades, a
great progress has been made in study of the QCD ground state and
a number of important results have been obtained that connect the
properties of the vacuum with the hadron characteristics treating the
QCD vacuum in the framework of the instanton liquid model
\cc{ILM, DP}. Considering QCD vacuum as an ensemble of instantons,
one can describe a number of the low-energy phenomena in strong
interactions on qualitative and quantitative levels \cc{REV,
DZK92}. The importance of the instanton induced effects in the
strong interaction is also supported by lattice simulations
\cc{REV, LAT}. The instanton picture is generally considered as a
fruitful and perspective framework for hadronic physics. The role
of instantons in the hard hadronic processes has been studied
intensively, both theoretically and experimentally. The contact
with the perturbative QCD results becomes possible providing the
unique information about the quark-gluon distribution functions in
the QCD vacuum and hadrons at low energy normalization point
\cc{D00, D03WF, DB03Corr}. The perspectives for an unambiguous
experimental detection of instanton effects are believed to be
optimistic and promising.

Although the QCD vacuum is known to play an important role in the
high-energy scattering processes, the direct investigation of these effects
remains a difficult task. The idea that the nontrivial vacuum
structure could be relevant in high energy hadronic processes was
explicitly formulated in the context of the soft Pomeron
problem in \cc{LOW, NUS, LN}, and further developed using the
eikonal approximation and the Wilson integral formalism in
\cc{NACH,KRP}.

In the present work, we report our recent results on investigation
of the instanton induced effects in high energy regime for the
electromagnetic (EM) quark form factor, {\it i.e.,} the amplitude
of the quark elastic scattering in an external color singlet gauge
field. Then, we investigate the role of instantons in the
diffractive quark-quark scattering and formation of the soft
Pomeron. All the considered cases manifest a similar structure and
are studied within the unified framework---the Wilson integral
approach, which allows to study both perturbative and
nonperturbative effects on the same ground. The method of
path-ordered Wilson integrals is known as a powerful (and
sometimes unique) tool in QCD which reformulates the theory in
terms of the gauge invariant quantities---the Wilson loops---
while the gauge fields are considered as chiral fields in the
space of all possible loops \cc{MMP}. The Green functions,
amplitudes, and cross sections can be expressed completely in
terms of the Wilson integrals over contours with geometry
determined by specific kinematics in an intrinsically
non-diagrammatic, ({\it i.e.,} nonperturbative) fashion \cc{WREN,
WILS, STEF3}.

\section{Electromagnetic quark form factor}

The behavior of the hadronic form factors in various energy
domains is one of the most important questions in the strong
interaction phenomenology.
The quark form factor has important phenomenological applications
since it enters the cross sections of a number of high energy hadronic processes
\cc{HARD}. For example, the total cross section of the Drell-Yan
process (normalized to the deep inelastic one) is determined by
the ratio of the time-like and space-like form factors \cc{PAR,
MAG}: \be {\s^{DY}_n \over \s^{PM}_n} \sim \Bigg|{F_q(Q^2) \over
F_q(-Q^2)}\Bigg|^2 \ , \ee where $\s^{PM}_n$ is the $n$-th moment
of the cross section calculated within the parton model.

The electromagnetic quark form factors are determined via the
elastic scattering amplitude of a quark in an external EM field:
\be {\cal M}_\m =F_q\[(p_1-p_2)^2\] \bar u(p_1) \g_\m v(p_2)  -
G_q
\[(p_1-p_2)^2\] \bar u(p_1) {\s_{\m\n} (p_1-p_2)_\n \over 2m}
v(p_2)
 \ \ , \label{eq:ampl}
\ee where $u(p_1), \ v(p_2)$ are the spinors of outgoing and
incoming quarks, and $\s_{\m\n} = [\g_\m, \g_\n]/2$. The
kinematics of the process is described by two invariants (see
Fig. 1a): \be s = (p_1 + p_2)^2 = 2m^2 (1 + \ch \c) \ \ , \ \ t =
- Q^2 =  (p_1 -p_2 )^2 = 2m^2 (1 - \ch \c) \ \ , \label{m2} \ee
where $\chi$ is the scattering angle. In
this work we assume that both the momentum transfer $- t$ and the
total center-of-mass energy $s$ are large compared to the quark
mass: \be (p_1p_2) \gg p_{1,2}^2 =m^2 \ \ , \ \ \hbox{or} \ \ \ch
\c \gg 1 \ . \ee In this regime the Pauli form factor $G_q$ is
power suppressed and will be neglected. However, it should be
emphasized that in low and moderate energy domains it becomes
important and there arise interesting perturbative and
nonperturbative effects (see, {\it e.g.,} recent works \cc{ET,
KOCH}).

The EM quark form factor is one of the simplest and convenient
objects for investigation of the double logarithmic behaviour of
the amplitudes in QCD in the high energy regime. From the
methodological point of view, it requires a perturbative
resummation procedure beyond the standard renormalization group
techniques. Besides this, the resummation methods developed for
this particular case can be applied to many other processes which
possess the logarithmic enhancements near the kinematic
boundaries. Similar resummation approach is also used in the study
of the near-forward quark-quark scattering and the evaluation of
the soft Pomeron properties \cc{KRP}. In the latter case, the
nonleading logarithmic terms become quite important.

The first example of large logarithm resummation was given in QED
by V.V. Sudakov for the case of an {\it off}-shell fermion in the
external Abelian gauge field in the double-logarithmic
approximation (DLA), where the terms of order of $(\a^n \ln^{2n}
Q^2)$ are taken into account while the contributions from $O(\a^n
\ln^{2n-1} Q^2)$ are neglected. The exponentiation of the leading
double logarithmic result was found in \cc{SUD}. In DLA the
exponentiated form factor behaves as a rapidly decreasing at high
momentum transfer function. This means that the elastic scattering
of a quark by a virtual photon is strongly suppressed. The
exponentiation for the {\it on}-shell form factor in the Abelian
case was obtained in the DLA in \cc{ONLLA} (the analytical
calculation of the full electron vertex in two loops was performed
in \cc{REM}).

These results eventually have been generalized for a more
complicated non-Abelian gauge theory. First, the resumed DLA
corrections in the QCD perturbative series were found to be
consistent with the exponentiation in \cc{LLANA} (the inelastic
{\it on-}shell form factor with emission of one and two gluons was
calculated in the same context in \cc{LLANAIE}; the role of the
quark Sudakov form factor in the description of $e^+e^-$
one-photon annihilation in quarks and gluons was considered in DLA
in \cc{FEL}), and the all-order non-Abelian exponentiation has
been proved in \cc{LLAAOR}.

The question if the non-leading logarithmic terms could upset the
DLA behaviour required a further work \cc{COLL}. The
all-(logarithmic)-order resummation was performed in the Abelian
case and the exponentiation was demonstrated in \cc{ALLLOGA}. In
\cc{ALLLOGNA} the non-Abelian all-order exponentiation for the
so-called hard part of the {\it on-}shell form factor has been
shown within the powerful factorization approach\footnote{Note,
that in this work the case where a time-like photon with large
invariant mass decays into a quark-antiquark pair was considered,
however it can be easily shown that the results remain true as
well in the case of a quark scattering in an external EM field.}.
However, the status of the soft part, containing all the infrared
(IR) and collinear singularities and, as a consequence, all
possible nonperturbative effects, remained unclear. The important
results on the IR properties of the QCD quark form factors was
obtained in \cc{KRC1} within the Wilson loop approach. In these
works, the soft part of the form factor has been presented as the
vacuum averaged exponent of the path integral of a gauge
field over the contour of a special form---an angle with sides of
semi-infinite length (Fig. 1a). The use of the gauge and renormalization
groups invariance allowed to derive the perturbative evolution
equation describing the high energy behaviour of the form factor
taking into account all (power unsuppressed) parts of the
factorized amplitude, both for the {\it on-} \cc{KRON} and {\it
off-}shell \cc{KROFF} cases. It was shown that the leading
asymptotics is controlled by the cusp anomalous dimension which
arises due to the multiplicative renormalization of the soft part,
and can be calculated within the Wilson integral formalism up to
the two-loop order \cc{KRC1}. Note that within the Wilson integral
approach, the non-Abelian exponentiation can be proved
independently \cc{NAEXP}, what is an important advantage of this
framework.

The efficiency of the Wilson integral approach has been
successfully demonstrated in a series of works \cc{STEF3,
STEF2}. In these papers the non-diagrammatic framework was
developed that allows to calculate the fermionic Green's
functions, Sudakov form factors, amplitudes and cross sections in
QED and QCD completely in terms of the world-line integrals, and
thus avoid complicated diagrammatic factorization analysis.

The results presented above allow one to conclude that the leading
high energy behaviour of the quark form factor in a non-Abelian
gauge theory is completely determined by the perturbative
evolution equation, and is given by the fast decreasing exponent:
\be \sim \exp\[- {2C_F \over \b_0}\cdot \ln Q^2 \ \ln \ln Q^2 +
O(\ln Q^2)\]\ , \label{eq:LAS} \ee where
$$C_F =\frac{N_c^2 -1}{2N_c},\ \ \ \ \ \b_0 = {11 N_c -2 n_f \over 3}\ .$$
This rapid fall off is not
changed by any other logarithmic contributions \cc{ALLLOGNA, KRON,
KROFF, STEF2}. However, the non-leading logarithmic corrections
are nevertheless important for evaluation of the numerical value
of the form factor at moderately large values of momentum
transfer. Some of them are of a purely perturbative origin ({\it
e.g.}, sub-leading logarithmic terms), while the others can be
attributed to the nonperturbative phenomena.

In the present work, we try to advocate the point of view that the
true nonperturbative effects can be taken into account
consistently in the evolution equation, and therefore they yield
the non-vanishing subleading (perhaps, parameterically suppressed,
but still logarithmic) contributions $\sim \ln \ Q^2$ to the high
energy behaviour of the amplitudes. Further, we analyze another
possible source of contributions which can be considered as
``nonperturbative''---the IR renormalon ambiguities (there are
plenty of papers on this subject, for the most recent reviews see
\cc{REN}). We demonstrate explicitly that they produce the
corrections with different IR structure compared to that one
generated by instantons. Moreover, as we show below these
renormalon effects disappear in the dimensional regularization
\cc{MAG} and in the analytical perturbation theory \cc{SHIR}, what
means that they could be merely treated as artifacts of the
incompleteness of the perturbative series resummation procedure.

In the following Sections we describe the consequences of the RG
invariance of the factorized form factor and derive the linear
evolution equation considering the nonperturbative input as the
initial value for the perturbative evolution. We show that in the
dilute phase the all-order single instanton contribution is
exponentiated in small parameter of instanton density. Then, these
nonperturbative effects are estimated in the weak-field
approximation within the instanton model of QCD vacuum. By using the
Gaussian simulation of the instanton profile function we show that
the weak field result approximates well the
all-order multi-instanton contribution. The
large-$Q^2$ behaviour of the form factor is analyzed taking into
account the leading perturbative and instanton induced
contributions. The consequences of the IR renormalon ambiguities
of the perturbative series and their relevance within the context
of some analytization procedures are also studied. Finally, the
Wilson integral techniques is applied to evaluation of the
instanton contributions to the high-energy behavior of the
$qq$-scattering amplitude in the Regge regime, that is the soft
Pomeron problem.

\section{Evolution equation for the quark form factor}

The classification of the diagrams with respect to the momenta
carried by their internal lines allows one to express the form
factor $F_q$ in the amplitude (\ref{eq:ampl}) in the factorized
form \cc{ALLLOGA, ALLLOGNA, KRON}
\be
F_q(q^2) = F_H(q^2/\m^2, \a_s)\cdot F_S(q^2/m^2, \m^2/\l^2,
\a_s) \cdot F_J(\m^2/\l^2, \a_s)\
 , \label{eq:fact1}
\ee where the hard, soft, and collinear (jet) part are separated
by scale parameters $\mu^2$, $\lambda^2$. It is assumed the
following relations between these scales $q^2>>\mu^2>>\l^2>1$.
Note, that in the present paper all dimensional variables are
assumed to be expressed in units of the QCD scale $\L_{QCD}$, so
that $q^2 = Q^2/\L_{QCD}^2$, {\it etc}. The arbitrary scale $\m^2$
is assumed to be equal to the UV normalization point.

Within the eikonal approximation the resummation of all
logarithmic terms coming from the soft gluon subprocesses allows
us to express $F_S$ in terms of the vacuum average of the gauge
invariant path ordered Wilson integral \cc{MMP, Pol} \be F_S
(q^2/m^2, \m^2/\l^2, \a_s) = W (C_\c; \m^2/\l^2, \a_s)  = {1
\over N_c} \Tr \Big<0\Big|   \pa \exp \left\{ i g \int_{C_\c}
\! d x_{\m} \hat A_{\m} (x) \right\}\Big|0\Big\> \ .
\label{1a} \ee In Eq. (\ref{1a}) the integration path
corresponding to the considered process goes along the closed
contour $C_\c$: the angle (cusp) with infinite sides (Fig. 1). We
parameterize the integration path $C_\c=\{z_\mu(t);
t=[-\infty,\infty]\}$ as follows \be z_{\mu}(t)=\left\{
\begin{array}
[c]{c}%
v_{1}t,\qquad-\infty<t<0,\\
v_{2}t,\qquad0<t<\infty.
\end{array}
\right. \label{path}\ee
The gauge field $ \hat A_{\m} (x) = T^a A^a_{\m}(x)\  \   \
(\Tr[T^a T^b] = \frac{1}{2}\delta^{ab}) \ $ belongs to the Lie
algebra of the gauge group $SU(N_c)$, while the Wilson loop
operator $\pa \exp\({ig\int\! dx A(x)}\)$ lies in its
fundamental representation.

The Wilson integral (\ref{1a}) is multiplicatively renormalizable
\cc{KRC1, WREN}: \be W(C_\c;
\m^2/\l^2, \a_s(\m^2)) = Z_{cusp} (C_\c; \M^2/\m^2, \a_s(\m^2))
\cdot W_{bare} (C_\c; \M^2/\l^2, \a_s(\mu^2)) \ , \label{eq:renw1} \ee
where $\M^2$ is the UV cutoff, $\m^2$ is the normalization
point, and $\l^2$ is the IR cutoff.
The presence of the IR
divergence in (\ref{eq:renw1}) is a common feature of {\it
on}-shell amplitudes in massless QCD.
Therefore, we can define the cusp anomalous dimension
$\G_{cusp}$:
\begin{eqnarray}
{1 \over 2}\ \G_{cusp} (C_\c; \a_s(\m^2)) &=& - \m^2 {d \over d
\m^2} \ln W(C_\c; \m^2/\l^2, \a_s(\m^2))=\label{eq:renw2}\\
&=& - \m^2 {d \over d
\m^2} \ln Z_{cusp}(C_\c; \M^2/\m^2, \a_s(\m^2) ) \ .
\nonumber
\end{eqnarray}
It can be shown that the cusp anomalous dimension
(\ref{eq:renw2}) is linear in the scattering angle $\c$ to all
orders of perturbation theory in the large-$q^2$ regime
\cc{KRC1}: \be \G_{cusp} (C_\c; \a_s) = \ln q^2 \ \G_{cusp}
(\a_s) + O(\ln^0 q^2)\ , \label{eq:cad} \ee where \be\G_{cusp}
(\a_s) = \sum_0^\infty \(\frac{\a_s}{\pi}\)^n C_n A_n \ , \ee
$C_n$ are maximally non-Abelian color factors ($C_n = C_F
N_c^{n-1}$ in lowest orders),  and $A_n$ are some numerical
factors. The observation (\ref{eq:cad}) is crucial for the
efficiency of the analysis of high energy behavior within the
Wilson integral formalism, since calculating the cusp anomalous
dimension only in leading (one-loop) order one obtains the
dominant contribution to the high energy asymptotics.

The total form factor $F_q$ is renormalization invariant
quantity and satisfies to the renorm group (RG) equation:
\be \m^2 {d \over d \m^2} \ F_q (\m^2, \a_s(\m^2))=
\(\m^2 {\pd \over \pd \m^2} + \b(\a_s){\pd \over \pd \a_s}\) \
F_q (\m^2, \a_s(\m^2)) =0 \ , \ee
which in the large-$q^2$ regime
leads to the following relations \be \m^2 {d \over d \m^2}\[
{\pd \ln F_H \over \pd \ln q^2}\] = - \m^2 {d \over d \m^2}
\[{\pd \ln F_S \over \pd \ln q^2} \]=
{1 \over 2}\ \G_{cusp} (\a_s) \ . \label{eq:anom} \ee
To avoid the problems with additional light-cone
singularities in the soft part, we work with logarithmic
derivatives in $q^2$ \cc{KRON, KRLC}. The collinear
part $F_J$ being independent on $q^2$ does not contribute to
these equations.

Then, from the Eqs. (\ref{eq:anom}, \ref{eq:renw2},
\ref{eq:cad}) one finds after simple calculations \cc{KRON}: \be
{\pd \ln F_H (q^2) \over \pd \ln q^2} = \int_{q^2}^{\m^2} \!
{d\x \over 2\x} \G_{cusp} (\a_s(\x)) + \G(\a_s(q^2)) \ ,
\label{eq:solH} \ee \be {\pd \ln F_S (q^2) \over \pd \ln q^2} =
- \int_{\l^2}^{\m^2} \! {d\x \over 2\x} \G_{cusp} (\a_s(\x)) +
{\pd \ln W_{np} (q^2) \over \pd \ln q^2} \ , \label{eq:solS} \ee
where the ``integration constant'' of the hard part reads \be
\G(\a_s(q^2)) = {\pd \ln F_H (q^2)\over \pd \ln q^2}\Bigg|_{\m^2
= q^2} \ , \label{eq:intch} \ee and $W_{np}$ arises as the
initial value of the soft part: \be {\pd \ln W_{np} (q^2) \over
\pd \ln q^2}  = {\pd \ln F_S(q^2) \over \pd \ln q^2}
\Bigg|_{\m^2 = \l^2} \ , \label{eq:intcs} \ee and is the only
quantity where, according to our suggestion, the nonperturbative
effects take place \cc{DCH1, DCH2}.

Then we get the $q^2$-dependence of the total form factor at large
$q^2$:
$$
\ln {F_q(q^2) \over F_q(q^2_0)} = $$ \be = -
\int_{q_0^2}^{q^2}\! {d\x \over 2\x} \ \[ \ln{q^2 \over \x} \
\G_{cusp}(\a_s(\x) ) - 2 \G(\a_s(\x))\] - \ln {q^2 \over q_0^2}\
\ \int_{\l^2}^{q_0^2} \! {d\x \over 2\x} \ \G_{cusp}(\a_s(\x) )
+  \ln {W_{np} (q^2) \over W_{np} (q_0^2)} \ . \label{eq:gen1}
\ee In the next subsection, we explicitly calculate the
perturbative quantities entering Eq. (\ref{eq:gen1}) in the one
loop perturbative QCD approximation and in the other parts we concentrate on
the nonperturbative part within ILM.

\section{Perturbative contributions to the Wilson integral}

The analysis of the hard contributions \cc{ALLLOGNA, KRON} at
large $q^2$ yields: \be {\pd \ln F_H(q^2/\m^2, \a_s) \over \pd
\ln q^2} = - {\a_s \over 2 \pi} C_F \(\ln {q^2 \over \m^2} - {3
\over 2} \) + O(\a_s^2)\ . \label{eq:hard1} \ee
For the hard ``integration constant''
(\ref{eq:intch}) one has: \be \G(\a_s(q^2)) = {3 \over 4}
{\a_s(q^2) \over \pi} C_F \ . \ee The expression
(\ref{eq:hard1}) is IR-safe, while the low-energy information is
accumulated in the soft part of the quark form factor $F_S$. The
Wilson integral (\ref{1a}) can be presented as a series:
\begin{eqnarray}
W(C_\c) && = 1 + {1 \over N_c}\Tr\<0| \sum_{n=2} \ (ig)^n
\int_{C_\c}\int_{C_\c} ...\int_{C_\c}
 \! dx_{\m_n}^n \ dx_{\m_{n-1}}^{n-1}... dx_{\m_1}^1 \cdot\nonumber\\ &&
\cdot \ \theta (x^n, x^{n-1}, ... , x^1) \  \hat A_{\m_n}
(x^n) \hat A_{\m_{n-1}} (x^{n-1})... \hat A_{\m_{1}} (x^1)|0\> \ ,
\label{expan1} \end{eqnarray} where the
function $\theta (x)$ orders the color matrices along
the integration contour.

The leading order of the expansion (\ref{expan1}) is given by
expression (see Fig. 1b-c): $$ \w1_{bare} (C_\c) = -\frac{g^2}{2}
\  \frac{1}{N_c} \Tr\(T^a T^b\) \ \int_{C_\c}\! dx_\m
\int_{C_\c}\! dy_\n \ {\cal D}_{\m\n}^{ab} (x-y) $$ \be =
-\frac{g^2}{2} \frac{1}{2 N_c} \ \int_{C_\c}\! dx_\m \int_{C_\c}\!
dy_\n \ {\cal D}_{\m\n} (x-y) \ , \label{g1} \ee where the gauge
field propagator ${\cal D}_{\m\n}(z)$ in $n$-dimensional
space-time $(n = 4 - 2 \ve)$ reads: \be {\cal D}_{\m\n}(z) =
\Big\<0\Big|{\cal T}A^{\bar a}_\m(z)A^{\bar a}_\n(0)\Big|0\Big\> \
, \ee The exponentiation theorem for non-Abelian path-ordered
Wilson integrals \cc{NAEXP} allows us to express (to one-loop
accuracy) the Wilson integral (\ref{1a}) as the exponentiated
one-loop term of the series (\ref{expan1}): \be W_{bare}(C_\c;
\ve, \M^2/\l^2) = \exp\[\w1_{bare}(C_\c; \ve, \M^2/\l^2) +
O(\a_s^2)\] \ . \label{eq:expn1} \ee In general, the expression
(\ref{g1}) contains UV and IR divergences, that can be
multiplicatively renormalized in a consistent way \cc{WREN}. In
the present work, we use the dimensional regularization for the UV
singularities, and define the ``gluon mass'' $\l^2$ as the IR
regulator.

The dimensionally regularized free propagator in
covariant gauge reads  $(n=4-2\ve) \ , \ \ve > 0$:
$${\cal D}_{\m\n} (z; \x)  = \(N_c^2 -1\) \prop(z;\x)\ , $$
\be \prop(z; \x) =  \m^{4-n} {1 \over i} \int \! \dk \ex^{-ikz}
\({g_{\m\n} \over k^2 -\l^2 +i0  } - \x {k_\m k_\n/\[k^2 -(1-\x)
\l^2 +i0\] \over k^2 -\l^2 +i0} \) \ , \label{Dpert}\ee where $\x$
is a gauge fixing parameter. It is convenient to use the
representation \be \prop(z) = g_{\m\n} \pd_z^2 \D_1(\ve, z^2,
\M^2/\l^2) - \pd_\m\pd_\n \D_2(\ve, z^2, \M^2/\l^2) \ ,
\label{st1} \ee where $\M^2$ is a parameter of dimensional
regularization.

By using integrals (\ref{A1}) - (\ref{A6}) from Appendix A, the dimensionally regularized
formula for the leading order (LO) term (\ref{g1}) can be written
as \cc{DCH1}: \be \w1_{bare} (C_\c; \ve, \M^2/\l^2, \a_s) = 8 \pi
\a_s C_F \ h (\c) (1 - \ve)  \D_1(\ve, 0, \M^2/\l^2) \ ,
\label{pe1} \ee where $h(\c)$ is the universal cusp factor: \be
h(\c) = \c \coth \c -1 \  ,\ee which at large-$q^2$ is given by:
\be \lim_{\chi\to\infty}{h(\c)}\to \chi \propto \ln {q^2 \over
m^2} \ . \label{eq:lar1} \ee In Eq. (\ref{pe1}), for the
perturbative gauge field one has \be \D_1 (\ve, 0, \M^2/\l^2) = -
{1 \over 16\pi^2} \(4\pi {\M^2 \over \l^2}\)^{\ve} \ {\G(\ve)
\over 1 - \ve} \ .  \label{eq:pr01} \ee  The independence of
the expression (\ref{pe1}) of the function $\D_2$ is a direct
consequence of the gauge invariance.

Thus, in
the one-loop approximation one gets \be W_{bare}(C_\c; \ve, \M^2/\l^2,
\a_s) = 1 - {\a_s \over 2\pi} C_F h(\c) \({1 \over \ve} - \g_E +
\ln 4\pi + \ln{\M^2\over\l^2} \), \ee and the cusp dependent
renormalization constant, within the $\overline{MS}$-scheme
which fixes the UV normalization point, reads: \be Z_{cusp}
(C_\c; \ve, \M^2/\m^2, \a_s(\m^2)) = 1 + {\a_s(\m^2) \over 2\pi}
C_F h(\c) \({1 \over \ve} - \g_E + \ln 4\pi \) +O(\a_s^2)\ . \ee

Using the Eq. (\ref{pe1}), one finds the known one-loop result
for the perturbative field, which  contains the dependence on
the UV normalization point $\m^2$ and IR cutoff $\l^2$ (see {\it
e.g.}, \cc{KRC1, STEF2}): \be \w1_{pt} (C_\c; \m^2/\l^2,
\a_s(\m^2)) =  - {\a_s (\m^2) \over 2 \pi} C_F  h(\c) \ln {\m^2
\over \l^2} + O(\a_s^2) \ . \label{5} \ee Therefore, in the
leading order the kinematic dependence of the expression
(\ref{g1}) is factorized into the cusp factor $h(\c)$ \cite{Pol}.

From the one-loop result (\ref{5}), the cusp anomalous dimension
which satisfies the RG equation (\ref{eq:renw2}) in one-loop
order is given by
\footnote{The cusp anomalous dimension is known up to two-loops,
see \cite{KRC1}.}: \be {\G_{cusp}^{(1)}} ( \a_s (\m^2)) = {\a_s
(\m^2) \over \pi} C_F \ . \label{cp} \ee

Substituting into the
Eq. (\ref{eq:gen1}) the anomalous dimension (\ref{cp}) with the
strong coupling constant given in the one-loop approximation,
one finds
$$ F_q^{(1)}(q^2) = $$
\be \exp\[- {2C_F \over \b_0}
\[\ln q^2 \(\ln{\ln q^2 \over \ln \l^2} -1 \)
- {3 \over 2} \ln {\ln q^2 \over \ln q_0^2 } + \ln q_0^2 \(1 -
\ln{\ln q_0^2 \over \ln \l^2} \)  \] +  W_{np}(q^2) \]
F^{(1)}(q_0^2)\ . \label{npc0} \ee Note, that the exponent in
Eq. (\ref{npc0}) has an unphysical singularity at $\l^2 =1$ (in
dimensional notations, $\bar\l^2 = \L_{QCD}^2$), {\it i. e., }
where the one-loop coupling constant $\a_s (\bar\l^2)$ has the
Landau pole. This feature can be treated in terms of IR
renormalon ambiguities (see next Section), and is often
considered as a signal of nonperturbative physics. In the
present paper, we will consistently separate the sources of
nonperturbative effects which can be attributed to uncertainties
of resummation of the perturbative series from the ``true''
nonperturbative phenomena. An important example of the latter is
provided by instanton induced effects within the instanton model
of QCD vacuum, which is considered in the Section 6.

\section{IR renormalon induced effects}

As it was pointed out at the end of the previous Section, the
perturbative evolution equation (\ref{npc0}) possesses an
unphysical singularity at the point $\l^2 = 1$. Therefore, it is
instructive to study the consequences of this feature. It is known
that the presence of the Landau pole in the one-loop expression
for the coupling constant leads to the IR renormalons \cc{REN}
resulting in power suppressed corrections. In the present
situation one can expect the corrections proportional to the
powers of both scales: $\m^2$ and $\l^2$. We will treat here the
power $\m^2$-terms to be strongly suppressed in large-$q^2$
regime, and focus on the power $\l^2$-corrections. To find them,
let us consider the perturbative function $\D_1(\ve, 0,
\M^2/\l^2)$ in the Eq. (\ref{pe1}). The insertion of the fermion
bubble 1-chain to the one-loop order expression (\ref{g1}) is
equivalent to replacement of the frozen coupling constant $g^2$ by
the running one $g^2 \to g^2 (k^2) = 4\pi \a_s(k^2)$ \cc{KRREN}
(for convenience, we work here in Euclidean space.): \be
\widetilde\D_1(\ve, 0, \M^2/\l^2) = - 4\pi \M^{2\ve} \int \! \dk
\a_s(k^2){\ex^{ikz} \delta(z^2)
 \over k^2(k^2+\l^2)}\ .
\label{ren1} \ee By using the integral representation for the
one-loop running coupling $\a_s(k^2) = \int_0^\infty \! d\s
(1/k^2)^{\s b }$, $b = \b_0 /4\pi$, we find: \be
\widetilde\D_1(\ve, 0, \M^2/\l^2) = - {1 \over \b_0(1-\ve)} \(4\pi
{\M^2 \over \l^2}\)^\ve \int_0^{\infty} \! dx\ {\G(1-x -\ve)
\G(1+x+\ve) \over (x +\ve ) \G(1 - \ve )} \({1 \over \l^2}\)^x \ .
\label{gamma} \ee To define properly the integral in r. h. s. of
Eq.(\ref{gamma}), one needs to specify a prescription to go around
the poles, which are at the points $\bar x_n = n , \ n \in
{\mathbb N}$. Thus, the result of integration depends on this
prescription giving an ambiguity proportional to $\(1 / \l^2 \)^n$
for each pole. Then, the IR renormalons produce the power
corrections to the one-loop perturbative result, which we assume
to exponentiate with the latter \cc{KRREN}. Extracting from
(\ref{gamma}) the UV singular part in vicinity of the origin $x
=0$, we divide the integration interval $[0, \infty]$ in two
parts: $[0, \d]$ and $[\d, \infty]$, where $\d < 1$. This
procedure allows us to evaluate separately the ultraviolet and the
renormalon-induced pieces. For the ultraviolet piece, we apply the
expansion of the integrand in $\D_1$ in powers of small $x$ and
replace the ratio of $\G$-functions by $\exp(-\g_E \ve)$: \be
\widetilde\D_1^{UV} (\ve, 0, \M^2/\l^2) = -  {1 \over \b_0 (1-
\ve)} \sum_{k, n=0} (-)^n { \(  \ln 4\pi - \g_E + \ln {\M^2  \over
\l^2}\)^k \over k! \ve^{n-k+1}}  \ \int_0^{\d} \! dx \ x^{n} \ \(1
\over \l^2 \)^x \ , \label{rnm1} \ee which after subtraction of
the poles in the $\overline{MS}$-scheme becomes: \be
\widetilde\D_1^{UV} (0, \m^2/\l^2) =  {1 \over \b_0 ( 1- \ve)}\
\sum_{n=1}\(\ln {\m^2 \over \l^2} \)^n {(-)^n \over n!} \
\int_0^{\d} \! dx x^{n-1} \(1 \over \l^2 \)^x \ .
\label{eq:iks}\ee In analogy with results of \cite{Mikh98}, this
expression may be rewritten in a closed form as \be
\widetilde\D_1^{UV} ( 0, \m^2/\l^2) = {1 \over \b_0 (1- \ve)}
\int_{0}^{\d}\frac{dx}{x} \[\ex^{-x \ln \m^2} -  \ex^{-x\ln \l^2}
\]. \label{Dex}\ee Then, using the relation \be {\pd  \w1 (q^2)
\over \pd \ln q^2}= 2 C_F (1-\ve) \widetilde \D_1^{UV}(0,
\m^2/\l^2) \ , \label{Dex1}\ee one finds \be \(\m^2 {\pd \over \pd
\m^2} + \b(g) {\pd \over \pd g} \) {\pd W^{(1)} (q^2) \over \pd
\ln q^2}
 = - {1 \over 2} \G_{cusp}^{(1)} (\a_s(\m^2)) \(1-\exp{\[-\d\frac{4\pi}{\b_0\a_s(\m^2)}\]}\).
\ee The exponent in the last equation yields the power
suppressed terms $\(1/q^2\)^\d$ in large-$q^2$ regime. In LLA
the Eq. (\ref{Dex1}) is reduced to: \be {\pd \w1 (q^2) \over \pd
\ln q^2} = - {2C_F \over \b_0} \( \ln{\ln \m^2 \over \ln \l^2}
\) \ . \label{ptr2} \ee The last expression obviously satisfies
the perturbative evolution equation (\ref{npc0}).

The remaining integral in Eq. (\ref{gamma}) over the interval
$[\d, \infty]$ is evaluated at $\ve =0$ since there are no UV
singularities. The resulting expression does not depend on the
normalization point $\m$, and thus it is determined by the IR
region including nonperturbative effects. It contains the
renormalon ambiguities due to different prescriptions in going
around the poles $\bar x_n$ in the Borel plane which yields the
power corrections to the quark form factor.

After the substitution $\m^2 = q^2$ and integration, we find in
LLA (for comparison, see Eq. (\ref{npc0})): \be F_q^{ren}(q^2) =
\exp\[- {2C_F \over \b_0} \ln q^2 \(\ln \ln q^2  - 1 \) -  \ln q^2
\Phi_{ren} (\l^2) \] F^{ren}(q_0^2)\ , \label{npc} \ee where the
function $ \Phi_{ren}(\l^2) = \sum_{k=0} \f_k (1/\l^2)^k$
accumulates the effects of the IR renormalons. The coefficients
$\f_k$ cannot be calculated in perturbation theory and are often
treated as ``the minimal set'' of nonperturbative parameters.  It
is worth noting that the logarithmic $q^2$-dependence of the
renormalon corrections in the large-$q^2$ regime is factorized,
and thus the Eq. (\ref{npc}) corresponds to the structure of
nonperturbative contributions found in the one-loop evolution
equation (\ref{npc0}). On the other hand, it can be shown that
careful account of partially resumed perturbative series yields,
sometimes, cancellation of the leading power corrections
associated with the leading renormalon contributions \cc{KRREN}.
As the corresponding nonperturbative terms are calculated
independently ({\it e.g.}, by means of the ILM) their direct
relation to the IR renormalon ambiguities should be questioned. In
our point of view, it allows us to separate the true
nonperturbative ({\it e.g.,} instanton induced, but not only)
effects from that ones related to ambiguities of the resumed
perturbative series.

The latter conclusion can be illustrated by considering the
consequences of an analytization (AN) of the strong coupling constant
\cc{SHIR} in the perturbative evolution equation. In this
approach, the one-loop strong coupling constant $\a_s(\m^2)$ is
replaced by the expression which is analytical at $\m^2 = 1$
({\it i.e.,} at $\L^2_{QCD}$ in dimensional variables): \be
\a_s^{AN}(\m^2) = {4\pi \over \b_0} \( {1 \over \ln \m^2} + {1
\over 1 - \m^2} \) \ . \label{eq:an1} \ee The direct
substitution of (\ref{eq:an1}) into the evolution equation
(\ref{eq:gen1}) yields (for brevity, we assume $q_0^2 = \l^2$):
\begin{eqnarray}
&& - \({\b_0 \over 2C_F} \) \ln F_q^{AN} (q^2)
= \ln q^2 \left(\ln{\ln q^2 \over \ln \l^2} - 1\right) -
{3 \over 2} \ln {\ln q^2 \over \ln \l^2 } +  \\
&& + \left(\ln q^2+{3\over2} \right)\(\ln {q^2 (\l^2-1)\over (q^2 -1)\l^2}
\) - {1 \over 2} \( \ln^2 q^2 - \ln^2 \l^2 \) -
\hbox{Li}_2 (1-q^2) + \hbox{Li}_2 (1-\l^2)  \ .
\nonumber\end{eqnarray} The functions $\hbox{Li}_2$ in the resulting
expression accumulate the power corrections of $q^2$ and IR scale
$\l^2$, but does not exhibit a singularity at $\l^2=1$. Therefore,
it gives no room for IR renormalons ambiguities, at least in the
considered approximation. Nevertheless, the power corrections of
nonperturbative origin do contribute to the large-$q^2$ behaviour.
Note, that the consequences of the analytization of the strong
coupling constant in the IR region have been studied also in
\cc{STAN}.

Another possible way to avoid the Landau pole at the integration
path was developed within the dimensional regularization (DR) scheme
\cc{MAG}. In this case, the running coupling reads \be \a_s^{DR}
(\ve; \m^2) =  {4 \pi \ve \over \b_0 \[ \(q^2 \)^\ve -1 \]} \ ,
\label{eq:dr1} \ee and for complex $\ve \ ,\  \hbox{Re}\  \ve <
0$, it has the Landau pole at the complex value of $\m^2$, thus
this singularity moves out of the integration contour. In the
limit $\ve \to 0$, the form factor reads \cc{MAG} (for
comparison, see Eq. (\ref{npc0}):
$$ F_q^{DR} (q^2) = $$ \be \exp\[- {2
C_F \over \b_0} \({\zeta(2) \over \ve} + \ln \ve \(\ln q^2 - {3
\over 2}\) + \ln q^2 \(\ln \ln q^2 -1 \) - {3 \over 2} \ln \ln
q^2 \) + O(\ve, \ve \ln \ve) \] \ . \ee This expression also
leaves no room for any renormalon effects. At the same
time, the nonperturbative contributions still take place since
they enter into the ``integration constant'' $W_{np}$,
(\ref{eq:solS}), which is not directly related to the analytical
properties of the coupling constant.

\section{Instanton induced contribution to the Wilson integral}

In this Section we consider explicitly the nonperturbative effects
within the instanton model of QCD vacuum. The instanton effects in
the high energy QCD processes has been actively studied since the
seventies \cc{EL, BAL}. Recently, the investigation of these
effects was renewed with promising perspectives \cc{KOCH, DCH1,
DCH2, Sh, KKL, RW, SRP, RING,  D03, CH1, DKZ92}. The Wilson
integral formalism is considered as a useful and convenient tool
in the instanton applications mainly due to significant
simplification in the path integral calculations if an explicit
form of the gauge field is known. Another important feature of
this approach is the possibility to make a correct analytical
continuation of the results obtained in the Euclidean space (where
the instantons are only determined) to the physical Minkowski
space-time where the scattering processes actually take place.
Namely, one maps the scattering angle, $\c$, to the Euclidean
space angle, $\g$, by analytical continuation \cc{EUC}
\be \c \to i\g \ , \label{MtoE}\ee
and performs the inverse transformation to the
Minkowski space-time in the final expressions in order to restore
the $q^2$-dependence.

Let us consider the instanton  contribution to the
function $W_{np}(q^2)$ from Eq.(\ref{eq:intcs}).
The instanton field has general form
\be \hat A_\m (x; \r) = A^a_{\m}
(x; \r) {\s^a \over 2} = {1 \over g}
 {\mathbb R}^{ab} \s^a {\eta^\pm}^b_{\m\n} (x-z_0)_\n \vf
(x-z_0; \r) , \label{if1}\ee where $\varphi_{I}(x)$ is the gauge
dependent profile function, ${\mathbb R}^{ab}$ is the color
orientation matrix, $\s^a$'s are the Pauli matrices,
${\eta^\pm}^a_{\m\n}=\varepsilon_{4a\m\n}\mp(1/2)\varepsilon_{abc}\varepsilon_{bc\m\n}$
are 't Hooft symbols, and $(\pm)$ corresponds to the instanton or
antiinstanton solution \footnote{Below we always consider only
topologically neutral instanton (I) and antiinstanton (A)
configurations, but since there are no differences in
contributions of I and A in the processes considered both
solutions will be called as instanton.}.

The averaging of the Wilson operator over the nonperturbative
vacuum is performed by the integration over the coordinate of the
instanton center $z_0$, the color orientation and the instanton
size $\r$.  The measure for the averaging over the instanton
ensemble reads $dI = d{\mathbb R} \ d^4 z_0 \ dn(\r) $, where $
d{\mathbb R}$ refers to the averaging over color orientation, and
$dn(\r)$ depends on the choice of the instanton size distribution.
Taking into account (\ref{if1}), we write the Wilson integral
(\ref{1a}) with contour (\ref{path}) in the single instanton
approximation in the form:
\begin{eqnarray}
w_I(\c) &=& {1\over N_c} \Tr \<0|
\exp \[ i \sigma^a
\(\hat n^a_1 \alpha(v_1, z_0) + \hat n^a_2 \alpha(v_2, z_0)\)\]|0\> = \label{wI1}\\
&=& {1\over N_c} \Tr \<0|
\cos{(\alpha(v_1, z_0))} \cos{(\alpha(v_2, z_0))}\({\mathbb I}\times {\mathbb I}\)-\nonumber\\
&-&\frac{1}{3}\(\hat n_1\cdot\hat n_2\)\sin{(\alpha(v_1, z_0))} \sin{(\alpha(v_2, z_0))}
\(\sigma^a\times\sigma^a\)
|0\> ,
\nonumber\end{eqnarray}
where $(i=1,2)$
\be \hat n^a_i  = \frac{(-1)^i}{s(v_i,z_{0})}
{\eta^\pm}^a_{\m\n} v_i^\mu z_0^\nu
 \ ,
\label{iin} \ee
\begin{eqnarray}
\alpha (v_i, z_0) &=&   s(v_i,z_{0}) \int_0^\infty \! d\l \vf\[(
v_i\l +(-1)^iz_0)^2; \r_c \] \ , \label{it2}
\end{eqnarray}
and
\begin{equation}
 s^2(v_i,z_{0})=z_0^2   -(v_iz_0)^2,
\label{sv} \end{equation} where $(v_1v_2) = \cosh \c$ in Minkowski
geometry. We omit the path ordering operator $\pa$ in (\ref{wI1})
because the instanton field (\ref{if1}) is a hedgehog in color
space, and so it locks the color orientation by space coordinates.
Let us note that due to
nonperturbative factor $g^{-1}$ in the instanton field (\ref{if1})
the Wilson integral (\ref{wI1}) is independent on the coupling constant.

At this stage we need to average over all possible ways of embedding $SU_c(2)$ into
$SU_c(3)$ by using the correspondence relations \cite{SVZ80}
\begin{eqnarray}
\langle \({\mathbb I}\times {\mathbb I}\)_{SU(2)}\>_{SU(3)}&=&
\frac{4}{9}\[ \({\mathbb I}\times {\mathbb I}\)_{SU(3)}+\frac{3}{32}\(\l^A\times\l^A\)_{SU(3)}\],
\label{SU23}\\
\langle \left( \s^a\times\s^a\)_{SU(2)}\>_{SU(3)}&=& \frac{3}{8}\(\l^A\times\l^A\)_{SU(3)},
\nonumber\end{eqnarray}
where $\l^A$ are generators of $SU(3)$ algebra $(A=1,2,...,8)$. Then we get
\begin{eqnarray}
w_I(\c) &=& {1\over 3} \Tr \Big<0\Big|
\frac{4}{9}\cos{(\alpha(v_1, z_0))} \cos{(\alpha(v_2,
z_0))}\({\mathbb I}\times {\mathbb I}\)
+\label{wI3}\\
&+&\frac{1}{8}
\[\frac{1}{3}\cos{(\alpha(v_1, z_0))} \cos{(\alpha(v_2, z_0))}
-\sin{(\alpha(v_1, z_0))} \sin{(\alpha(v_2, z_0))}\hat n_1^a\hat
n_2^a\]\(\l^A\times\l^A\) \Big|0\Big> \ . \nonumber\end{eqnarray}
The resulting gauge invariant contribution to the Wilson loop of
the single instanton taken in all orders in gauge field becomes
\cc{DCH1}
\begin{eqnarray}
w_I(\chi) &=&  \frac{2}{3}\int dn(\rho) \[w_c^I(\chi)+w_s^I(\chi) - w_c^I(0)-w_s^I(0)\],\label{it11} \\
w_c^I(\chi)&=&
\int \! d^4 z_0 \ \cos \ \alpha (v_1, z_0) \cos \ \alpha (v_2, z_0)\ , \label{it1c}\\
w_s^I(\chi)&=& - \int \! d^4 z_0 \ (\hat n^a_1\hat n^a_2)\sin \
\alpha (v_1, z_0) \sin \ \alpha (v_2, z_0) \ ,
\nonumber\end{eqnarray}
where the normalized color correlation factor is
\be
\hat n^a_1\hat n^a_2 =
-\frac{ \eta _{\mu \nu }^{a }v_{1}^{\mu}z_{0}^\nu \eta _{\rho \sigma }^{a}v_{2}^{\rho }z_{0}^\sigma }
{s(v_1,z_{0})s(v_2,z_{0})}. \label{n1n2}\ee
The Eq. (\ref{it11}) takes into
account the subtraction of the infinite self-energy parts of the quark form
factor. Thus, to calculate the instanton contribution one need to
consider only the vertex corrections.

\section{Exponentiation of the instanton contributions in the dilute regime}

On the basis of the exponentiation theorem \cc{NAEXP} for the
non-Abelian path-ordered exponentials it is well known that
perturbative corrections to the Sudakov form factor are
exponentiated to high orders in the QCD coupling constant. The
theorem states that the contour average $W_{P}(C)$ can be
expressed as \be W_{P}(C)=\exp{\[
\sum_{n=1}^{\infty}\(\frac{\a_s}{\pi}\)^n\sum_{W\in W(n)}
C_n(W)F_n(W) \]}, \label{Exp}\ee where summation in the
exponential is over all diagrams $W$ of the set $W(n)$ of the
two-particle irreducible contour averages of $n$th order of the
perturbative expansion. The coefficients $C_n(W)\propto C_F
N_c^{n-1}$ are the ``maximally non-Abelian'' parts of the color
factor corresponding to the contribution coming from a diagram $W$
to the total expression (\ref{Exp}) in the contour
gauge\footnote{The contour gauge is defined as a gauge where the
condition $P\exp \left(ig\int_{x_0}^x dz_\mu \hat A_\mu
(z)\right)=1$ is fulfilled \cc{KIR}.}, and the factor $F_n(W)$ is
the contour integral presented in the expression for $W$. This
means that the essential diagrams are only those, which do not
contain the lower-order contributions as subgraphs and, as a
result, the higher-order terms are non-Abelian.

Let us now demonstrate how the single instanton contribution is
exponentiated in the small instanton density parameter, treating
the instanton vacuum as a dilute medium \cite{CDG78}. The gauge
field is taken to be the sum of individual instanton fields in
the singular gauge, with their centers at the points $z_{j}$'s.
In this gauge, the instanton fields  fall off rapidly at
infinity, so the instantons may be considered individually in
their effect on the Wilson loop. Moreover, the contribution of
infinitely distant parts of the contour may be neglected and
only those instantons will influence the loop integral, which
occupy regions of space-time intersecting with the quark
trajectories. Since the parameterization of the loop integral
along rays of the angle plays the role of the proper time, a
time-ordered series of instantons arises and has an effect on
the Wilson loop. Thus, the contribution of $n$ instantons to the
loop integral $W_{I}(\chi)$ can be written in the dilute
approximation as \be W^{(n)}_I(\chi)=\Tr \( U^1
U^2...U^{n}U^{n\dagger}...U^{2\dagger }U^{1\dagger }\) , \ee
where the ordered line integrals $U_{i}$'s
\begin{eqnarray*}
U^{j}(\chi) &=&T\left\{ \exp{\(ig\int_0^{\infty}\! d\s\ v_1^\m
A_\m(v_1\s-z_j)\)} \exp{\(ig\int^0_{-\infty} \! d\t \ v_2^\m
A_\m(v_2\t-z_j)\)} \right\}
\end{eqnarray*}
are associated with individual instantons with the positions
$z_{j}$'s. Because of the wide separation of the instantons in
the dilute phase and rapid fall off of fields in the singular
gauge, the upper and lower limits of the line integrals are
extended to infinity. The line integrals $U^{i\dagger}$'s take
into account the infinitely distant part of the contour
that goes from
$+\infty$ back to $-\infty$ and in the singular gauge they are
$U^{i\dagger}=1$. For $U^{j}(U^{j\dagger })$, the integral is
taken over the increasing (decreasing) time piece of the loop.

Then, the expression is further simplified by averaging over the gauge
orientations of instantons. The averaging is reduced to
substitution of $U^{j}$ by $g_{j}U^{j}g_{j}^{-1}$, where $g_{j}$
is an element of colour group, and independent integration of
each $g_{j}$ over the properly normalized group measure is
performed. Under this averaging one gets \be U^{n}U^{n\dagger
}\rightarrow w_I^{(n)}(\chi)=\frac{1}{N_{c}} \Tr \( U^{n}U^{n\dagger }\), \ee
which is just the single instanton contribution
as it is given by Eqs. (\ref{it11}). If the
averaging is done in the inverse order, from $n$ down to $1$,
the entire loop integral becomes the product of traces \be
W^{(n)}_I(\chi)\rightarrow\lim_{n\to \infty }\prod_{j=1}^{n}
w_I^{(j)}(\chi). \ee Since the individual instantons are
considered to be decoupled in the dilute medium, the total
multiple instanton contribution to the vacuum average of the
Wilson operator simply exponentiates the all-order single
instanton term $w_I(\chi)$ in (\ref{it11}), and one has
\begin{eqnarray}
W_I(\chi)&=&\lim_{n \to \infty }\left\{ 1+\frac{1}{n} w_I(\chi)
\right\} ^{n} =\exp [w_I(\chi)]. \label{WDYinst}
\end{eqnarray}

Thus, we prove that in the dilute regime the full instanton
contribution to the quark form factor is given by the exponent
of the all-order single instanton result (see Fig. 1e). The
exponentiation arises due to taking into account the
multi-instanton configurations. As it is well known, in
QED there occurs the exponentiation of the one-loop result due
to Abelian character of the theory. In the instanton case, the
analogous result arises since instantons belong to the
$SU(2)$ subgroup of the $SU(3)$ color group and the path-ordered
exponents coincide with the ordinary ones.

The following comments are in order. First, the nonperturbative
exponentiated expressions are strictly correct only as long as
the instanton density $n(\rho)$ is small. Second, it is supposed
that $U(z_{0})$ is evaluated using the singular gauge form of
$A_{\mu }^{inst}$. On the other hand, $\Tr\left( UU^{\dagger }\right) $
is identically the ordered loop integral
for a single instanton and is gauge invariant. It is therefore
legal to use the nonsingular gauge of $A_{\mu }^{inst}$ in
evaluating the trace (a more handing gauge for computation).

\section{Large-$q^2$ behaviour of the instanton  contribution in the weak field approximation}

Let us first consider the weak-field approximation to the
instanton contribution to the quark form factor. In this limit the
leading instanton induced term in Eq. (\ref{it11}) is given by
general expression (\ref{g1}) (Fig. \ref{fig0}c) with the corresponding
dimensionally regularized instanton correlator \be {\cal
D}^I_{\mu\nu}(z)= \l^{n-4} \ \dI \
  \int \! \dk \
\tilde A^a_\m (k; \r) \tilde A^a_\n (-k; \r) \ex^{-ikz} \ .
\label{i1} \ee By using the Fourier transform of the instanton
field embedded into the $SU(3)$ color group \be \tilde A^a_\m (k;
\r) = - {2i \over g } {\mathbb R}^{a\a}{\eta^\pm}^\a_{\m\s} k_\s
\tilde \vf'(k^2; \r ) \ , \label{Fourier}\ee where $\tilde \vf
(k^2; \r)$ is the Fourier transform of the instanton profile
function $\vf (z^2; \r)$ and $\tilde \vf'(k^2; \r)$ is its
derivative with respect to $k^2$, and the properties of the color
rotation matrices and 't Hooft symbols
$${\mathbb R}^{a\b} {\mathbb R}^{a\g} = \d^{\b\g} \ , \
\eta^{\a}_{\m\r}\eta^{\a}_{\n\l} = g_{\m\n}g_{\r\l} -
g_{\m\l}g_{\n\r} + \varepsilon_{\m\n\r\l} \ , $$ one can express
the instanton correlator (\ref{i1}) in the form similar to Eq.
(\ref{st1}) \be {\cal D}^I_{\mu\nu}(z) = \frac{1}{g^2}\ \(g_{\m\n}
\pd^2 - \pd_\m\pd_\n \) \ \D_I(z^2) \ ,  \ee where \be \D_I(z^2) =
- \l^{n-4} \ \int \! dn(\r) \ \dk \ \ex^{-ikz}
\[2\tilde \vf'(k^2; \r ) \]^2  \ . \ee Then, applying the same
method as described in perturbative case, one gets the LO
instanton contribution in the form \be w_I^{(LO)}(\c) =
\frac{1}{N_c} \ h(\c) \int\! dn(\r) \ \D_I(0, \r^2\l^2) \ ,
\label{wILO}\ee where we use the same IR cutoff $\l^2$, while the
UV divergences do not appear at all due to the finite instanton
size. In the singular gauge with the profile function given by \be
\varphi_{I}(x)=\frac{\r^{2}}{x^{2}\( x^2+\r^2\) },
\label{Inst_Profile} \ee one gets: \be \D_I (0, \r^2\l^2) = {\pi^2
\r^4 \over 4} \[\ln (\r^2 \l^2) \ \F_0(\r^2\l^2) + \F_1(\r^2\l^2)
\]\ ,
\label{DIinst}\ee where \be \F_0 (\r^2\l^2) = {1 \over \r^4\l^4} \int_0^1\!
{dz \over z (1-z)} \ \[1+ \ex^{\r^2\l^2} - 2 \ex^{z \cdot
\r^2\l^2} \] \ \ , \ \ \lim_{\l^2\to 0}\F_0(\r^2\l^2) = 1 \ ,
\ee and \be \F_1(\r^2\l^2) = \sum_{n=1}^\infty\int_0^1 \! dxdydz
\ {[-\r^2\l^2 (xz + y(1-z))]^n \over n! n}\ex^{\r^2\l^2 [xz +
y(1-z)] } \ \ , \ \ \lim_{\l^2\to 0}\F_1(\r^2\l^2) = 0 \ \ee are
the IR-finite expressions.

At high energy the instanton  contribution is reduced to the form:
\be {\pd  \ln W_I(q^2) \over \pd  \ln q^2} = {\pi^2 \over 4N_c}
\dI \ \r^4 \[\ln (\r^2 \l^2) \ \F_0(\r^2\l^2) + \F_1(\r^2\l^2) \]
\equiv - B_I(\l^2)\ .  \label{II1}\ee Here we used the
exponentiation of the single-instanton result in a dilute
instanton ensemble (see \cc{DCH1} and the previous Section) and
took only the LO term of the weak-field expansion (\ref{g1}): $\w1
= w_I + (higher \ order \ terms)$.

In order to estimate the magnitude of the instanton effect we take
the instanton size distribution found by 't Hooft \cc{tH} and
multiply it by the exponential suppressing factor (which was
suggested in \cc{DEMM99} in the framework of constrained instanton
model and assumed in \cc{SH2} in order to describe the lattice
data \cc{LAT}) : \be dn(\r) = {d\r \over \r^5} \ C_{N_c} \[2\pi
\over \a_s(\m_r) \]^{2N_c} \exp\[- {2\pi \over \a_s (\m_r)}
\] \(\r\m_r\)^{\b} \exp\(- 2 \pi \s \r^2\) \ , \label{dist1}
\ee where the constant $C_{N_c} = 4.6/\pi^2 \ \exp(-1.679
N_c)/\[(N_c-1)! (N_c-2)!\] \approx 0.0015 $, $\s\approx (0.44 \ GeV)^2$ is the string
tension,  $\b= \b_0+O(\a_s(\m_r))$, and $\m_r$ is the
normalization point \cc{MOR}.

Given the distribution (\ref{dist1}) the main parameters of the
instanton liquid model, the instanton density $n_c$ and the mean
instanton size $ \r_c $, will read: \be n_c = \int_0^\infty \!
dn(\r) = {C_{N_{c}} \G (\b/2 - 2) \over 2} \[2\pi \over \a_S(\bar
\r^{-1}) \]^{2N_c}
\[{\L_{QCD} \over \sqrt{2\pi \s}}\]^{\b} (2\pi \s)^2 \ ,
\ee \be \r_c = {\int_0^\infty \! \r \ dn(\r) \over \int_0^\infty
\! dn(\r)} = {\G(\b/2 - 3/2) \over \G(\b/2 - 2)} {1 \over \sqrt{2
\pi \s} } \ . \label{nbar} \ee In Eq. (\ref{nbar}) we choose, for
convenience, the normalization scale $\m_r$ of order of the
instanton inverse mean size $\r^{-1}_c$, taking into account that
the distribution function (\ref{dist1}) is the RG-invariant
quantity up to $O(\a_s^2)$ terms \cc{MOR}. Note, that these
quantities correspond to the mean size $\r_c$ and density $n_c$ of
instantons used in the model \cc{ILM}, where the size distribution
(\ref{dist1}) is approximated by the delta-function: \be
 dn(\r) = n_c \d(\r-\r_c) d\r \ .
\label{dn}\ee

Thus, we find the leading instanton contribution (\ref{II1}) in
the form: \be B_I(\lambda^2)=  \frac{K f}{2N_c} \ln{2\pi \s \over \l^2} \[1 +
O\({\l^2\over 2\pi\s}\)\] , \label{pow1} \ee where \be K =
{\G(\b_0/2) [\G (\b_0/2-2)]^3 \over 2 \ [\G(\b_0/2-3/2)]^4}
\approx 0.74 \ , \ee and we used the one loop expression for the
running coupling constant
\be \a_s ( \r_c^{-1}) = - {2\pi \over\b_0 \ln \ { \r_c \L}}\ . \ee
In (\ref{pow1}) we introduces the packing fraction parameter
\be f=\pi^2  n_c{ \rho_c}^4
\ee
that characterizes diluteness of the instanton liquid.
Within the conventional picture its value is estimated to be
\be f\approx 0.12 \ ,\ee
if one takes the model parameters as (see \cc{REV}):
\be  { n_c} \approx 1 fm^{-4}, \ \ { \r_c} \approx 1/3 fm . \label{param} \ee

Therefore, from Eqs. (\ref{II1}) and (\ref{npc}), we find the
expression for the quark form factor at large-$q^2$ with the
one-loop perturbative contribution and the nonperturbative
contribution (the function $W_{np}$ in Eq. (\ref{npc0})) found in
the instanton model: \be F_q(q^2) = \exp\[- {2C_F \over \b_0} \ln
q^2 \ln \ln q^2  - \ln q^2 \ \(B_I - {2C_F \over \b_0} \) + O(\ln
\ln q^2) \] F_0 (q_0^2; \l^2) \ . \label{eq:final} \ee
Numerically the coefficient $B_I$ is small factor as compared to the perturbative term
\be
B_I   \approx 0.03 \ll \frac{2C_F}{\beta_0} \approx0.24, \
\label{BI}\ee
where to get this estimate we take the IR cutoff parameter $\l$
of order of $\L_{QCD}\approx 350$ MeV.
It is clear, that while the asymptotic ``double-logarithmic''
behaviour is controlled by the perturbative cusp anomalous
dimension, the leading instanton correction results in a finite
renormalization of the subleading perturbative term
(Fig.\ref{fig1}). Note, that the instanton correction has the
opposite sign compared to the perturbative logarithmic term.

It is also important to note that the results obtained are quite
sensitive to the way one makes the integration over instanton
sizes finite. For example, if one used the sharp cutoff then the
instanton would produce strongly suppressed power corrections
like $\propto (1/q)^{\beta_0}$. However, we think that the
distribution function (\ref{dist1}) should be considered as more
realistic, since it reflects more properly the structure of the
instanton ensemble modeling the QCD vacuum. Indeed, this shape
of distribution was recently advocated in \cc{SH2, DEMM99} and
supported by the lattice calculations \cc{LAT} (for comparison,
see, however, \cc{UKQCD, SCH2}).

\section{All-order calculations of the Wilson loop for Gaussian profile}

The weak field limit used in the previous Section may deviate from
the exact result. Here, we are going to test its accuracy
considering the properties of the single instanton contribution to
the Wilson loop, $w_I(\chi)$, defined in Eq. (\ref{it11}), which
contains contributions from the gauge field taken in all orders.

We have to note that in realistic instanton vacuum model there are
two essential effects: stabilization of the instanton density with
respect to unbounded expansion of instantons in size (see Eqs.
(\ref{dist1}) and (\ref{dn})), and screening of instantons by
surrounding background fields. To take into account these features
we approximate first the narrow instanton size distribution by the
$\delta$- function as in Eq. (\ref{dn}) and assume that the
integration over the instanton size is already performed. As it
was discussed above, the screening effect modifies the instanton
shape at large distances leading to the constrained instantons
\cite{DEMM99}. To take into account this screening and to have
also simpler analytical form for $w_I(\chi)$, we use in this
Section the Gaussian Anzatz for the instanton profile function
\begin{equation}
\varphi _{G}(x^{2})=\frac{1}{\rho_c^2}e^{-x^{2}/\rho_c ^{2}} \ .
\label{Inst_Profile_G}
\end{equation}
The parameters in this expression are fixed from the requirement
to reproduce the vacuum average value of the $\langle
A_\m^a(0)A_\n^a(0)\rangle$ which is finite for the instanton field
in the singular gauge and equals\footnote{This average is infinite
in the regular gauges. Thus, the singular gauge realizes the
minimal Landau gauge for the instantons.}: \be \langle
g^2A_\mu^a(0)A_\mu^a(0)\rangle = 12\pi^2\rho_c^2n_c \ . \ee Below
we equal the parameter $\rho_c$ to unity and restore the
dependence on it at the end of calculations.

The phases (\ref{it2}) corresponding to the profile (\ref{Inst_Profile_G})
are calculated as
\begin{equation}
\alpha \left( v_{1,2},z_{0}\right) =s(v_{1,2},z_{0})
\frac{\sqrt{\pi }}{2}e^{- s^{2}(v_{1,2},z_{0})} \mathrm{erfc}\[
\mp \left(v_{1,2}z_{0}\right)\]  \ ,  \label{PhaseG}
\end{equation}
where with definitions (\ref{path}) one has\footnote{In these
expressions and below we make formally analytical continuation to
the Minkowski space that is inverse to the transformation
(\ref{MtoE}).}
\begin{eqnarray}
\left( v_{1}z_{0}\right) &=&z_4,\qquad \left( v_{2}z_{0}\right)
=z_4\cosh \chi +iz_{3}\sinh \chi,
  \label{p1x0} \\
s^{2}(v_1,z_{0})&=&z_3^2+z_{\perp}^2,\ \ s^{2}(v_{2},z_{0})=
\left( z_{3}\cosh \chi -iz_4\sinh \chi \right) ^{2}+z_{\perp
}^{2}\ . \label{p2x0}
\end{eqnarray}
We take the reference frame where the scattering point is in the
origin and define the scattering vectors of the quark as
\begin{eqnarray}
v_1 &=&\( 1,0,0_{\perp }\) \ ,\qquad v_2=\( \cosh \c , i\sinh \c
,0_{\perp }\)\  ,  \label{p1p2}
\end{eqnarray}
$$
v_1^2=v_2^2=1 \ ,\qquad \( v_1 v_2\) =\cosh \c \ ,
 $$
where the velocities $v_1=p_1/m$ and $v_2=p_2/m$ determine the
scattering plane. In this system the instanton position is given
by $z_0=(z_4,z_3,z_\perp)$.

However, due to exponentially large oscillations at large $\c$
that occur during integration over instanton position, it is not
easy to use the closed form for the phases (\ref{it2}) in general
case and for the Gaussian profile (\ref{Inst_Profile_G}), in
particular. We have to note that these complications do not arise
in the case of high-energy elastic quark-quark scattering
considered in \cc{Sh}. In the latter case, the angle dependence
simply factorizes from the integrand in the asymptotic regime (see
the next Section).

In order to cancel exponentially large oscillations, we need to
integrate first over the instanton position $z_0$. For this
purpose, by using the explicit form (\ref{Inst_Profile_G}) we
expand the expressions (\ref{it1c}) in powers of the phases
(\ref{it2}) and change the order of integrations: \be
w_{c}^{I}\left( \chi \right) =\sum_{n=1}^{\infty
}\sum_{m=1}^{\infty}(-1)^{n+m} \frac{\left\langle \alpha
_{1}^{2n}\alpha _{2}^{2m}\right\rangle (\chi)}{\left( 2n\right)
!\left( 2m\right) !},\ \ w_{s}^{I}\left( \chi \right)
=\sum_{n=0}^{\infty }\sum_{m=0}^{\infty}(-1)^{n+m}
\frac{\left\langle \alpha _{1}^{2n+1}\alpha
_{2}^{2m+1}\right\rangle (\chi)} {\left( 2n+1\right) !\left(
2m+1\right) !}, \label{wc}\ee where \be \left\langle \alpha
_{1}^{2n}\alpha _{2}^{2m}\right\rangle(\chi) =\int_{0}^{\infty
}\prod_{i=1}^{2n}d\lambda _{i}\int_{0}^{\infty
}\prod_{j=1}^{2m}d\lambda _{j}^{\prime }\int
d^{4}z_{0}s^{2n}\left( v_{1},z_{0}\right) s^{2m}\left(
v_{2},z_{0}\right) e^{-\left[ \left( v_{1}\lambda
_{i}-z_{0}\right) ^{2}+\left( v_{2}\lambda _{j}^{\prime
}+z_{0}\right) ^{2}\right] } \ , \label{wc1}\ee and
\begin{eqnarray}
\left\langle \alpha _{1}^{2n+1}\alpha _{2}^{2m+1}\right\rangle
(\chi) &=&\int_{0}^{\infty }\prod_{i=0}^{2n+1}d\lambda
_{i}\int_{0}^{\infty }\prod_{j=0}^{2m+1}d\lambda _{j}^{\prime
}\int d^{4}z_{0}s^{2n}\left( v_{1},z_{0}\right) s^{2m}\left(
v_{2},z_{0}\right) s_{12}^{2}(z_{0})\cdot
\nonumber \\
&&\cdot e^{-\left[ \left( v_{1}\lambda _{i}-z_{0}\right)
^{2}+\left( v_{2}\lambda _{j}^{\prime }+z_{0}\right) ^{2}\right] }
\  . \label{ws1}\end{eqnarray} Note, that the integrands possess
the symmetry with respect to the change of variables \be
z_3\rightarrow z_3\cosh \chi + iz_4\sinh \chi \ , \qquad
z_4\rightarrow z_4\cosh \chi - iz_3\sinh \chi \ . \ee Above we
introduced the notation for the color spin correlation factor (its
normalized definition given above in (\ref{n1n2}))
\begin{eqnarray}
s_{12}^{2}(z_{0}) &=& z_{3}\left( z_{3}\cosh \chi -iz_4\sinh \chi
\right) +z_{\perp}^{2}\cosh \chi \ . \label{s12}\end{eqnarray}

Let us further make the change of variables $\lambda $ by introducing
the total and partial lengths
\be
\{\lambda _{i}\}_{N}\rightarrow \left\{ L=\sum_{i=1}^{N}\lambda _{i},x_{i}=%
\frac{\lambda _{i}}{L}\right\} ,\qquad \{\lambda _{i}^{\prime
}\}_{M}\rightarrow \left\{ L^{\prime }=\sum_{j=1}^{M}\lambda
_{j}^{\prime },y_{j}=\frac{\lambda _{i}^{\prime }}{L^{\prime }}
\right\} \ \ee with new measures given by \be \int_{0}^{\infty
}dLL^{N-1}\int d\{x\}_{N},\qquad \int_{0}^{\infty }dL^{\prime
}L^{\prime M-1}\int d\{y\}_{M} \ ,
 \ee
where $\int
d\{x\}_{N}=\int_{0}^{1}dx_{1}...\int_{0}^{1-x_{1}-...-x_{N-2}}dx_{N-1}$.

After these transformations we come to the general expression
for the phase averages (\ref{wc1}) and (\ref{ws1}) with arbitrary $M$ and $N$
\begin{equation}
\left\langle \alpha _{1}^{N}\alpha _{2}^{M}\right\rangle(\chi) =
\int_{0}^{\infty}dL\int_{0}^{\infty }dL^{\prime }
e^{-\frac{2}{N+M}LL^{\prime }\cosh \chi}G_{N,M}(L,L^\prime,\chi)
F_{N,M}(L)F_{M,N}(L^{\prime }) \ ,  \label{aNaM}
\end{equation}
with definitions
\begin{eqnarray}
&&F_{N,M}(L)=
 \\
&&=L^{N-1}\int d\{x\}_{N}\exp \left\{ - \frac{1}{N+M}\left[
L^{2}+(N+M-2)\sum_{i}^{N}\lambda _{i}^{2}- 4\sum_{i>j}^N\lambda
_{i}\lambda_{j}
 \right] \right\} \ ,
\nonumber\end{eqnarray}
\begin{equation}
G_{N,M}(L,L^\prime,\chi)=\int d^{4}z_{0}s^{2n}\left(
v_{1},z_{0}\right) s^{2m}\left( v_{2},z_{0}\right) s_{12}^{2\eta
}(z_{0})e^{-(N+M)\left( z_{\perp }^{2}+z_4^{\prime
2}+z_{3}^{\prime 2}\right) } \ ,  \label{G_NM}
\end{equation}
where $z_4^{\prime }=z_4-\frac{L-L^{\prime }\cosh \chi }{N+M}$ and $%
z_{3}^{\prime }=z_{3}+\frac{iL^{\prime }\sinh \chi }{N+M}$.
The definition for $F_{M,N}(L^{\prime })$ is similar to one of
$F_{N,M}(L)$, the power $\eta $ in $G_{N,M}(L,L^{\prime },\chi)$
is equal to one for $w_{s}^{I}\left(\chi \right) $ and to zero
for $w_{c}^{I}\left( \chi \right)$.

In principle, the integral in $G_{N,M}(L,L^{\prime },\chi)$ may be
done analytically for the Gaussian profile in arbitrary order of
expansion for any finite values of $\chi$. In this case, after the
$d^{4}z_{0}$ integration there remain no complex valued variables
anymore and the Euclidean (transverse) and Minkowski (longitudinal
in $\chi $) dependencies get factorized. But in practice only a
few first terms can be analyzed. Nevertheless, as we will see
below, in the large $\chi$ limit it is possible to make the
partial summation of the double sums in the expressions
(\ref{wc}). Let us demonstrate these statements in more detail.

Consider the lowest order one-loop and two-loop contributions given by (Fig. \ref{fig3})
\begin{equation}
w^I(\chi) = \frac{2n_c}{3} \left( -\langle
\alpha_1\alpha_2\rangle(\chi)|_S +\frac{1}{6}\langle
\alpha_1^3\alpha_2\rangle(\chi)|_S +\frac{1}{6}\langle
\alpha_1\alpha_2^3\rangle(\chi)|_S +\frac{1}{4}\langle
\alpha_1^2\alpha_2^2\rangle(\chi)|_S +...\right) \ ,
\label{FewOrder}\end{equation} where $\langle
\alpha_1^M\alpha_2^N\rangle(\chi)|_S=\langle
\alpha_1^M\alpha_2^N\rangle(\chi)- \langle
\alpha_1^M\alpha_2^N\rangle(0)$. The simplest one-loop diagram
(Fig. \ref{fig3}a) corresponds to the first term in
(\ref{FewOrder}) ( $N=M=1$ term in Eq. (\ref{aNaM})). In that
case, the integrals may be done analytically (by using the Eqs.
(\ref{Ias1}) and (\ref{Ias2})) and the final result reduces to the
weak field expression (\ref{wILO}) with the function $\Delta_1$
given by
\begin{equation}
\Delta_1^{G}(x^{2})=-\frac{\pi
^{2}\rho_c^4}{4}e^{-x^{2}/2\rho_c^2} \ . \label{Dx2_G}
\end{equation}
Thus, in the lowest order we have the result ($f=\pi
^{2}n_{c}\rho_c^4$)
\begin{eqnarray}
w_G^{(1,1)}\left(\chi \right) =-\frac{f}{12}h(\chi) \ ,
\label{w11}\end{eqnarray} and its small and large $\chi$ behavior
\begin{eqnarray}
w_G^{(1,1)}\left(\chi \to 0 \right) = -\frac{f}{36}\chi^2 \ ,\ \ \
\ \ w_G^{(1,1)}\left(\chi \to \infty\right) =-\frac{f}{12}\chi \ .
\label{w11L}\end{eqnarray} It is important to note, that due to
screening of instantons at large distances the result
(\ref{Dx2_G}) is IR-finite contrary to the case of the pure
instanton solution in (\ref{DIinst}).

If one neglected the color spin factors in (\ref{G_NM}), one would
get another dependence on the scattering angle
\begin{eqnarray}
w_{Gs}^{(1,1)}\left(\chi \right) &=&-\frac{f}{6}\left(1-\frac{  \chi}{\sinh \chi}\right) \ ,\\
w_{Gs}^{(1,1)}\left(\chi \to 0\right) &=&-\frac{f}{36}\chi^2 \ ,\
\ \ \ \ w_{Gs}^{(1,1)}\left(\chi \to\infty\right) =-\frac{f}{6} \
, \nonumber\end{eqnarray} thus getting the asymptotic which is not
enhanced by the large factor $\chi$.

The rest of Eq. (\ref{FewOrder}) corresponds to the two-loop
diagrams shown in Fig. (\ref{fig3}b) $(N=1,\ M=3$ in (\ref{aNaM}))
and in Fig. \ref{fig3}c  ($N=M=2$ in (\ref{aNaM})). Then, the
functions $F_{N,M}(L)$ and $G_{N,M}(L,L^{\prime },\chi)$ becomes
\begin{eqnarray}
F_{1,3}(L) &=&\exp{\(-\frac{3}{4}L^{2}\)} \ ,\qquad
F_{2,2}(L)=\sqrt{\frac{\pi}{2}}\exp{\(-\frac{1}{4}L^2\)}\hbox{erf}{\(\frac{L}{\sqrt{2}}\)}
\
,\qquad \nonumber\\
F_{3,1}(L) &=&L^{2}\int_{0}^{1}dx\int_{0}^{1-x}dx^{\prime
}e^{-\frac{1}{4}L^{2} \left[ 3+8(x^{2}+x^{\prime 2}+xx^{\prime
}-x-x^{\prime })\right] } \ , \label{F13}\end{eqnarray}
\begin{eqnarray}
&&G_{2,2}(L,L^\prime,\chi) =\int d^{4}z_{0}s^{2}\left(
v_{1},z_{0}\right) s^{2}\left( v_{2},z_{0}\right) e^{-4\left(
z_{\perp }^{2}+t^{\prime
2}+z_{3}^{\prime 2}\right) }= \label{G22}\\
&&=\frac{\pi^2}{2^{12}}
\[L^{\prime 2}\sinh^2\chi(L^2\sinh^2 \chi -6) - 8LL^\prime \sinh^2\chi\cosh\chi-6L^2\sinh^2\chi+52+8\cosh^2\chi
\] \ , \nonumber\\
&&G_{1,3}(L,L^\prime,\chi) =\int d^{4}z_{0}s^{2}\left(
v_{1},z_{0}\right)
s_{12}^{2}(z_{0})e^{-4\left( z_{\perp }^{2}+t^{\prime 2}+z_{3}^{\prime2}\right) }=\nonumber\\
&&=\frac{\pi^2}{2^{12}}
\[-LL^{\prime 3}\sinh^4 \chi +10 L^\prime \sinh^2\chi(L^\prime\cosh\chi+L)-60\cosh\chi
\] \ .
\nonumber\end{eqnarray} We compare the full result and the
one-loop and two-loop approximations to it at small $\chi $ in
Fig. \ref{f2}.

In Fig. \ref{f3} we also present the results corresponding to the calculations without color spin
factors, $s\left( v_i,z_{0}\right) $. In the latter case, the
coordinate integral $\int d^{4}z_0$ in (\ref{G_NM}) may be performed easily with the result
\begin{equation}
G^s_{N,M}(L,L^\prime,\chi)=\frac{\pi^2}{(M+N)^2} \ .
\label{GS_NM}
\end{equation}
The next order calculations may be done in a similar way. Finally,
it may be shown that in the limit of large scattering angle $\chi$ the asymptotics
reads $w_{Gs}^{I}\left( \chi \rightarrow \infty \right)
\sim \mbox{const} $. Thus, in this case one has weaker
asymptotics than the asymptotics with color spin factor
included.

In the following, we are interested in the limit $\chi \rightarrow
\infty $, where the coefficient of $\chi $ is free of the
light-cone singularities and therefore it has a well defined limit
for the on-shell quark momenta, $p_{1}^{2}=p_{2}^{2}=0$. To find
this asymptotics, we take into account that $F_{N,M}(L\rightarrow
0)\neq 0$ and $G_{N,M}(L,L^{\prime })$ are polynomials in $L$ and
$L^{\prime }$. With these properties and in the large $\chi$
limit, the $L$ and $L^{\prime }$ integrations may be performed
analytically
\begin{eqnarray}
&&\lim_{\chi \rightarrow \infty }\int_{0}^{\infty
}dL\int_{0}^{\infty }dL^{\prime }\left( LL^{\prime }\right)
^{n}e^{-\alpha LL^{\prime }\cosh \chi
}F_{N,M}(L)F_{M,N}(L^{\prime })=
\nonumber \\
&&= \frac{n!\chi }{\left( \alpha \cosh \chi\right)
^{n+1}}F_{N,M}(0)F_{M,N}(0) \ ,
\label{AsI0} \\
&&\lim_{\chi \rightarrow \infty }\int_{0}^{\infty
}dL\int_{0}^{\infty} dL^{\prime }\left( LL^{\prime }\right)
^{n}L^{m}e^{-\alpha LL^{\prime }\cosh \chi
}F_{N,M}(L)F_{M,N}(L^{\prime }) =\nonumber
\\
&&=\frac{n!}{\left( \alpha \cosh \chi \right)
^{n+1}}F_{M,N}(0)\int dLL^{m-1}F_{N,M}(L) \ .
\label{AsInt}\end{eqnarray} The important result is that the only
diagonal terms with equal powers of $L$ and $L'$ provide the
leading in $\chi$ asymptotics.

Let us consider the contribution to the asymptotics from the terms
with $m=0$ and arbitrary $n$ of $w_{s}^{I}\left( \chi \right) $
(Fig. \ref{fig6}a). This contribution is reduced to the element
\begin{eqnarray}
&&\left\langle \alpha _{1}^{2n+1}\alpha _{2}^{1}\right\rangle
=-\cosh \chi \int_{0}^{\infty }dLL^{2n}\int_{0}^{\infty
}dL^{\prime }\int
d\{x\}_{2n+1}e^{-L^{2}\sum_{i}^{2n}x_{i}^{2}-L^{\prime 2} }\cdot
\\
&&\cdot \int d^{4}z_{0} \left( z_{3}^{2} - iz_{3}z_4\tanh \chi
+z_{\perp}^{2}\right) \left( z_{3}^{2}+z_{\perp }^{2}\right) ^{n}
e^{-\left[2(n+1)z_{0}^2-2z_4\left( L-L^{\prime }\cosh \chi \right)
+ 2iz_{3}L^\prime\sinh \chi\right] } \ . \nonumber\end{eqnarray}
The $z_{0}$ integration gives (see Eqs. (\ref{GI1}) and
(\ref{GaussInt}))
\begin{eqnarray}
&&\left\langle \alpha _{1}^{2n+1}\alpha _{2}^{1}\right\rangle
=-\pi ^{2}\cosh \chi \int_{0}^{\infty }dLL^{2n}\int_{0}^{\infty
}dL^{\prime }\int
d\{x\}_{2n+1}e^{-L^{2}\sum_{i}x_{i}^{2}-L^{\prime 2}}
e^{\frac{L^{2}+L^{\prime 2}-2LL^{\prime }\cosh \chi
}{2(n+1)}}\cdot
\nonumber\\
&&\cdot\sum_{k=0}^{n}C_{k}^{n}\frac{k!}{\left[ 2(n+1)\right] ^{k+2}}\left[ \frac{k+1%
}{2(n+1)}\left( \frac{iL^{\prime }\sinh \chi }{2(n+1)}\right)
^{2(n-k)}\sum_{s=0}^{n-k}\overline{C}_{s}^{2(n-k)}y^{s}+\right.
\nonumber\\
&&\left. +\left( \frac{iL^{\prime }\sinh \chi }{2(n+1)}\right)
^{2(n-k+1)}\sum_{s=0}^{n-k+1}\overline{C}_{s}^{2(n-k+1)}y^{s}-\left( \frac{%
iL^{\prime }\sinh \chi }{2(n+1)}\right) ^{2(n-k+1)}\sum_{s=0}^{n-k}\overline{%
C}_{s}^{2(n-k)+1}y^{s}+\right.
\nonumber\\
&&\left. +\frac{iL\tanh \chi }{2(n+1)}\left( \frac{iL^{\prime }\sinh \chi }{%
2(n+1)}\right) ^{2(n-k)+1}\sum_{s=0}^{n-k}\overline{C}_{s}^{2(n-k)+1}y^{s}%
\right] \ , \label{A2n1A1}\end{eqnarray} where we introduce
notation
\begin{equation}
y=\frac{2(n+1)}{\left( iL^{\prime }\sinh \chi \right) ^{2}} \ .
\label{y}
\end{equation}
The Eq. (\ref{A2n1A1}) is still a general expression. To find its
large-$\chi $ limit, we need to analyze, according to
(\ref{AsI0}), the coefficients of the maximal powers of the
diagonal terms $\left( LL^{\prime }\right) ^{2n}$. In the first
and fourth terms in the brackets of (\ref{A2n1A1}), the only terms
with $k=s=0$ provide the leading $\chi $ asymptotics and the other
one gives the subleading contribution. The second and third terms
in the brackets give dominant asymptotics if $k+s\leq 1$. Their
sum provides the leading $\chi$-asymptotics, while the terms of
higher powers in $\chi \cosh \chi $ are canceled. Keeping only
leading terms in the Eq. (\ref{A2n1A1}), one gets
\begin{eqnarray}
&&\left. \left\langle \alpha _{1}^{2n+1}\alpha
_{2}^{1}\right\rangle \right| _{\chi \rightarrow
\infty}= -\frac{\pi ^{2}\cosh \chi
}{16(n+1)^3}\int_{0}^{\infty }dLL^{2n}\int_{0}^{\infty
}dL^{\prime } \int
d\{x\}_{2n+1}e^{-L^{2}\sum_{i}x_{i}^{2}-L^{\prime 2}}\cdot
\nonumber\\
&&\cdot e^{\frac{\left[ L^{2}+L^{\prime 2}-2LL^{\prime }\cosh
\chi \right] } {2(n+1)}}\left( \frac{iL^{\prime }\sinh \chi
}{2(n+1)}\right) ^{2n}\left(
2n+3-\frac{LL^{\prime }\sinh ^{2}(\chi )}{\left( n+1\right) \cosh (\chi )}%
\right) \ . \label{A2n1A1x}\end{eqnarray} As $\chi\to\infty$, all
integrals can be done analytically by using Eq. (\ref{AsI0}):
\begin{equation}
\left. \left\langle \alpha _{1}^{2n+1}\alpha _{2}^{1}\right\rangle
\right| _{\chi \rightarrow \infty }=\frac{(-1)^{n+1}\pi
^{2}(2n)!}{2^{4n+3}(n+1)^2} \chi \ . \label{A2n1A1As}
\end{equation}

Now, let us consider the contribution to the asymptotics
of $w_{c}^{I}\left(\chi\right)$ from
the terms with $m=1$ and arbitrary $n$ (Fig. \ref{fig6}b):
\begin{eqnarray}
&&\left\langle \alpha _{1}^{2n}\alpha _{2}^{2}\right\rangle
=\cosh ^{2}\chi \int_{0}^{\infty }dLL^{2n-1}\int_{0}^{\infty
}dL^{\prime }L^{\prime }\int
d\{x\}_{2n}\int_{0}^{1}dye^{-L^{2}\sum_{i}x_{i}^{2}-L^{\prime
2}\left( y^{2}+(1-y)^{2}\right) }\cdot
\nonumber\\
&&\cdot\int d^{4}z_{0}\left( \frac{z_{\perp }^{2}}{\cosh ^{2}\chi
}+z_{3}^{2}+2iz_{3}z_4\tanh \chi -z_4^{2}\tanh ^{2}\chi \right)\cdot
\nonumber\\
&&\cdot\sum_{k=0}^{n}C_{k}^{n}z_{\perp}^{2k}z_{3}^{2(n-k)}
e^{-\left[ 2(n+1)z_{0}^2+2z_4\left( L-L^{\prime }\cosh \chi
\right) +2iz_{3}\sinh \chi \right] } \ .
\end{eqnarray}
By using the table integrals Eq.(\ref{GI1}) and (\ref{GaussInt}),
the Gaussian integration over $z_{0}$  yields
\begin{eqnarray}
&&\left\langle \alpha _{1}^{2n}\alpha _{2}^{2}\right\rangle =
\pi ^{2} \int_{0}^{\infty }dL\int_{0}^{\infty }dL^{\prime
}\left( LL^{\prime }\right) L^{2(n-1)}\int
d\{x\}_{2n}\int_{0}^{1}dye^{-L^{2}\sum_{i}x_{i}^{2}-L^{\prime
2}\left( y^{2}+(1-y)^{2}\right) }\cdot
\nonumber\\
&&\cdot e^{\frac{L^{2}+L^{\prime 2}-2LL^{\prime }\cosh \chi
}{2(n+1)}} \sum_{k=0}^{n}C_{k}^{n}\frac{k!}{\left[ 2(n+1)\right]
^{n+3}} \left( \frac{L^{\prime 2}\sinh^2 \chi }{2(n+1)}\right)
^{(n-k)}\cdot
\nonumber\\
&&\cdot\left[ \sum_{s=0}^{n-k}\overline{C}_{s}^{2(n-k)}y^{s}
\left(-\frac{L^2\sinh^2\chi}{2(n+1)}+\frac{3}{2}+k+s\cosh^2\chi\right)-
\right.
\nonumber\\
&&\left.
-\sum_{s=0}^{n-k-1}\overline{C}_{s}^{2(n-k)}(n-k-s)y^{s+1}LL^\prime
\sinh^2\chi\cosh\chi \right] \ , \label{A2nA2}\end{eqnarray} with
$y$ given by Eq.(\ref{y}). Again, the large $\chi $ limit of this
expression stems from the diagonal terms $\left( LL^{\prime
}\right) ^{2n}$
\begin{eqnarray}
&&\left. \left\langle \alpha _{1}^{2n}\alpha
_{2}^{2}\right\rangle \right|
_{\chi \rightarrow \infty }=\nonumber\\
&& \frac{(-1)^{n-1}\pi ^{2}\cosh\chi }{[2(n+1)]^{2(n+2)}}
\int_{0}^{\infty }dL\int_{0}^{\infty }dL^{\prime } \left(
LL^{\prime }\sinh\chi\right)^{2n-1} \int
d\{x\}_{2n}\int_{0}^{1}dye^{-L^{2}\sum_{i}x_{i}^{2}-L^{\prime
2}\left( y^{2}+(1-y)^{2}\right) }\cdot
\nonumber\\
&&\cdot e^{\frac{L^{2}+L^{\prime 2}-2LL^{\prime }\cosh \chi
}{2(n+1)}} \left[ n(n+1)(2n-1)-4n(n+1)LL^\prime\sinh\chi
+(LL^\prime\sinh\chi)^2 \right] \ .
\end{eqnarray}
After integration one finds that the coefficient of the leading
asymptotics is equal to zero, and therefore
\begin{equation}
\left. \left\langle \alpha _{1}^{2n}\alpha _{2}^{2}\right\rangle
\right| _{\chi \rightarrow \infty }=\mbox{const} \ .
\end{equation}
Moreover, it is possible to show that further leading asymptotic
terms appear only if $n\geq3$ and $m\geq3$, but they are highly
suppressed numerically.

Thus from (\ref{it11}), (\ref{wc}) and (\ref{A2n1A1As}), we find
that the leading correction to the quark form factor is given by
\begin{equation}
w_{G}(q^{2})=-
\sum_{n=0}^\infty\frac{1}{16^n(n+1)^2(2n+1)}\cdot\frac{f}{12} \ln
q^2= -1.0053\frac{f}{12} \ \ln q^{2} \ , \label{SG2}\end{equation}
that means that the weak field limit (\ref{w11L}) is a good
approximation for the Gaussian profile function
(\ref{Inst_Profile_G}). This estimate of the logarithmic
coefficient---$B_G\approx 0.01$, is compatible with the estimate
(\ref{BI}) obtained in the single instanton approximation.

We have to comment that in the general (non-Gaussian) case the
weak-field limit may deviate from the exact result. Nevertheless,
it is reasonable to expect that by taking the instanton solution
in the singular gauge which concentrate the field at small
distances and allows us to prove the exponentiation theorem for
the Wilson loop in the instanton field \cc{DCH1}, one gets a good
numerical estimate of the full effect. Thus, the resulting
diminishing of the instanton contributions with respect to the
perturbative result appears to be reasonable output. The analysis
of all-order instanton contribution performed in the last part of
this Section for a Gaussian profile function shows that the weak
field approximation can be justified, but an additional
investigation of this problem is required.

\section{Instanton model of Pomeron}

Soft hadronic collisions are described successfully within the
Regge phenomenology, with the Pomeron exchange being dominating at
high energy. The Pomeron is considered as an effective exchange in
the $t$ channel by the object with vacuum quantum numbers and with
positive charge parity $C=+1$. That is why the idea that the
nontrivial structure of the QCD vacuum is relevant in describing
its mechanism. To illustrate this idea, let us consider the high
energy diffractive quark-quark scattering, where there is a hope
that for small momentum transfer the nonperturbative effects give
dominant contribution. One of the simplest models of the Pomeron
is based on the use of exchange by two nonperturbative gluons. The
nonperturbative part of the gluon propagator is given by (in the
Landau gauge) \be \Big<0\Big| :A^a_\m(x)A^b_\n(0):\Big|0\Big>
=\frac{1}{g^2}\frac{\d^{ab}}{N_{c}^{2}-1} \int
\frac{d^{4}k}{\(2\pi\) ^4}\ex^{-ikx}
\(g^{\m\n}-\frac{k^{\m}k^{\n}}{k^{2}}\)G_{np}(k^2) \ .
\label{GPland}\ee

In the Abelian gauge model considered originally by Landshoff and
Nachtmann \cite{LN}, the nonperturbative gluon propagator (without
a color factor) $G_{np}(k^{2})$ is related to the correlation
function describing the gauge invariant gluon field strength
correlator (nonlocal gluon condensate). In general non-Abelian
case this correlator has the form
$$
\Big<0\Big| :G_{\m\n}(x)\pa \exp \[ ig\int_{0}^{x}dz^{\a}A_\a
(z)\] G_{\r\s}(0):\Big| 0\Big> =
$$
$$
=\int \frac{d^{4}k}{\( 2\pi \) ^{4}}\ex^{-ikx}\[ \(
D_{0}(k^{2})+D_{1}(k^{2})\) k^{2}\( g_{\m\r}g_{\n\s}- g_{\mu
\sigma }g_{\nu \rho }\) +\right.
$$
\be
\left. +D_{1}(k^{2})\( k_\m k_\r g_{\n\s}-k_\m k_\s g_{\n\r}+
k_{\n}k_{\s} g_{\m\r} - k_{\n}k_{\r}g_{\m\s}\) \] \ , \label{GlFS}
\ee where the first tensor structure is called non-Abelian part
and the second one
is Abelian part. Indeed, in the Abelian gauge model without monopoles $%
D_{0}(k^{2})\equiv 0$, and $D_{1}(k^{2})=G_{np}(k^{2})$. It is
this property that has been used in \cite{LN} to relate the
Pomeron properties to the value of the gluon condensate.

However, in the non-Abelian model one has the opposite situation.
Really, for the QCD instantons we find \cite{DEM97,DEMM99}
$D_{1}(k^{2})\equiv 0$ and $D_{0}(k^{2})$ is nonzero. In the
realistic model of the QCD vacuum, where the interaction with
vacuum fields of large scale, $R$, is important, the instanton
ceases to be exact solution of the equations of motion, but the
so-called constrained instanton approximate solution (CI) can be
constructed \cite{DEMM99}. It has been shown that the constrained
instanton has exponentially decreasing asymptotics at large
distances ($\sim R$). The constrained instanton has topological
number $\pm 1$ like an instanton; however, it is not self-dual
field. Thus, in the realistic QCD, the small non-zero part of
$D_{1}(k^{2})$ appears. Very similar results have been found in
the lattice simulations of the gluon field strength correlator
\cite{DEL}.

Thus, within the non-Abelian models there is no direct
connection of the gluon propagator to the gluon field strength
correlator. So, let us explicitly consider the instanton part of
the gluon propagator. The Fourier transform of the instanton
field is defined as \be
\widetilde{A}_{\m}^a(k)=\frac{1}{g}\eta_{\m\n}^a k_{\n}\widetilde{\phi}%
(k^{2}) \ ,  \label{Agen,F} \ee where \be
\widetilde{\phi}(k^{2})=\frac{4\pi^2 i}{k^{2}}\int_{0}^{\infty
}dxx^{3}J_{2}(\left\vert k\right\vert x)\varphi (x^{2}) \ ,
\label{Phi,F} \ee and $J_{2}(z)$ is the Bessel function. The
function $\widetilde{\phi}(k^{2})$ is related to the function
$\widetilde{\varphi}(k^{2})$ defined in (\ref{Fourier}) by
\begin{equation}
\widetilde{\phi}(k^{2})=-2i\frac{\partial\widetilde{\varphi}(k^{2})}{\partial
k^2} \ .
\end{equation}
The explicit form of the Fourier transform of the pure instanton
solution is well known (in the singular gauge) \be \widetilde{\phi
}^{I}(k^{2})=i\frac{\left( 4\pi \right) ^{2}}{k^{4}}\left[ 1-
\frac{(\rho_c k)^{2}}{2}K_{2}\left( \rho_c k\right) \right] \ ,\ \
\ \widetilde{\phi }^{I}(k^{2})=\left\{
\begin{array}{c}
\frac{i\( 2\pi \) ^{2}\rho_c ^{2}}{k^{2}} \ ,\qquad
k^{2}\rightarrow 0,
\\
\frac{i\left( 4\pi \right) ^{2}}{k^{4}} \ ,\qquad k^{2}\rightarrow
\infty \ .
\end{array}%
\right. \label{Phi(p)_SingI_As}\end{equation}

The (constrained)
instanton profile may be chosen in the form ({\it c.f.}  Eq. (\ref{Inst_Profile}))
\begin{equation}
\varphi_{CI}(x)=\frac{\r^{2}_{CI}(x^2)}{x^{2}\(
x^2+\r^2_{CI}(x^2)\) },\ \ \ \ \ \rho_{CI}^{2}(x^{2}) =2\left(
\frac{\rho_c }{R}\right) ^{2}x^{2}K_{2}\left( \frac{\left|
x\right| }{R}\right), \label{CI}\ee where $K_2(z)$ is the modified
Bessel function. The constrained solution saves its form at short
distances, but changes it at large ones: \be \widetilde{\phi
}_{CI}(k^{2})=\left\{
\begin{array}{c}
\frac{i\pi ^{2}}{4}R^{4}I_{CI} \ ,\qquad k^{2}\rightarrow 0\ , \\
\frac{i\( 4\pi \)^{2}}{k^{4}} \ ,\qquad k^{2}\rightarrow \infty \ ,%
\end{array}%
\right.  \label{Phi(p)_SingCI_As} \ee where the constant $I_{CI}$
is given by \be I_{CI}=\int_{0}^{\infty }du\ u^{2}\varphi_{CI} (u
R^2)\ . \ee

Now the Fourier transform of the single instanton contribution to
the gluon propagator (\ref{GPland}) becomes
\begin{eqnarray}
G_{np}(k^2) =-4n_ck^{2}\widetilde{\phi}^{2}( k^2) \ .
\label{GProp_II}
\end{eqnarray}
Thus, we see that the nonperturbative gluon propagator (\ref{GPland}), (\ref{GProp_II}) and
gluon field strength correlator (\ref{GlFS}) are quite different functions, and
the relation between them that is valid in the Abelian gauge model
is destroyed in the non-Abelian case.

From (\ref{Phi(p)_SingI_As}) and (\ref{Phi(p)_SingCI_As})
one gets the asymptotics of the instanton part of the
gluon propagator
\begin{equation}
G^{I}(k^{2})=\left\{
\begin{array}{c}
\( 2\pi \) ^{4}n_{c}\rho_c ^{4}k^{-2}%
\ ,\qquad k^{2}\rightarrow 0 \\
\( 4\pi \) ^{4}n_{c}k^{-6}
\ ,\qquad k^{2}\rightarrow \infty%
\end{array}%
\right. \ ,\quad G^{CI}(k^{2})=\left\{
\begin{array}{c}
\frac{\pi ^{4}n_{c}R^{8}}{16}I_{CI}^{2}k^{2} \ ,\qquad
k^{2}\rightarrow 0 \\
\( 4\pi \) ^{4}n_{c}k^{-6} \
,\qquad k^{2}\rightarrow \infty%
\end{array}%
\right. \ . \label{GProp_As}\end{equation}

Calculating (in a very similar way as in the Landshoff-Nachtmann
model) the invariant $\mathcal{T-}$matrix element of the
quark-quark scattering at large energy, $s$, and small transferred
momentum, $t$, we get
\begin{equation}
\left\langle q_k(p_{3})q_l(p_{4})\left\vert \mathcal{T}\right\vert
q_m(p_{1})q_n(p_{2})\right\rangle \Big|_{s \to \infty} \rightarrow
iI(t)\quad \overline{u}(p_{3})\gamma ^{\mu
}u(p_{1})\overline{u}(p_{4})\gamma ^{\mu
}u(p_{2})\delta_{km}\delta_{ln} \ , \label{Tmatr}\end{equation}
with
\begin{equation}
I(t)=\frac{1}{72}\int \! \frac{d\vecc k_{\perp }}{(2\pi
)^{2}}G\left[ \left( \vecc k_{\perp }+\frac{1}{2}\vecc q_{\perp }
\right) ^{2}\right] G\left[ \left( \vecc k_{\perp }-\frac{1}{2}%
\vecc q_{\perp }\right) ^{2}\right]\  , \label{It}\end{equation}
where
$$
s=(p_1+p_2)^2=2m^2(1+\cosh\c),\ \
t=-(p_3-p_1)^2=-q_\perp^2,
$$
$G(k^2)$ is defined in (\ref{GProp_II}) with $k^2\to \vecc
k_{\perp }^2$. Except of the numerical coefficient, this
expression is in agreement with the Landshoff-Nachtmann formula.
This agreement is due to the specific features of the instanton
induced interaction. From the infrared behavior of the instanton
induced propagator (\ref{GProp_As}), it is clear that $I(0)$ is
infinite for the pure instanton solution (\ref{Phi(p)_SingI_As}),
but it is finite for the constrained instanton solution
(\ref{CI}). This fact, also noted recently in \cite{SH2}, was one
of the arguments to construct the constrained instanton solution.
The form of $I(t)$ is presented in Fig. \ref{fig8} for the
constrained instanton configuration.

It is possible to show that the result (\ref{Tmatr}) is the weak
field limit of the more general expression. Indeed, the
quark-quark scattering amplitude may be expressed in terms of the
vacuum average of the gauge invariant path ordered Wilson integral
\cite{NACH,KRP} \be T^{kl}_{mn}(s,t) = -2is\int d^2\vecc b_{\perp}
\ex^{i b_{\perp} q} W^{kl}_{mn}(\g,\vecc b^2_\perp) \ ,
\label{Tqq}\ee where the Wilson line function $W^{kl}_{mn}$ is
given by \be W^{kl}_{mn}(\g,b^2_\perp)=\Bigg<0\Bigg| \pa \exp
\left\{ i g \int_{C_{qq}} \! d x_{\m}\hat A_{\m}(x) \right\}
\Bigg|0\Bigg\>^{kl}_{mn} \ . \label{1ab} \ee In Eq. (\ref{1ab}),
the corresponding integration path goes along the closed contour
$C_{qq}$: two infinite lines separated by the transverse distance
$\vecc b_\perp$ and having relative scattering angle $\chi$. We
parameterize the integration path $C_\c=\{z_\mu(\lambda);
\lambda=[-\infty,\infty]\}$ as follows \be
z_{\mu}(\lambda)=\left\{
\begin{array}
[c]{c}%
v_{1}\lambda \ ,\qquad-\infty<\lambda<\infty \ ,\\
v_{2}\lambda + b_\perp \ ,\qquad-\infty<\lambda<\infty \ ,
\end{array}
\right. \ee with $\( v_1 v_2\) =\cosh \c$ and $\vecc b_\perp$
being the impact parameter.

By making the steps similar to the previous Section, one arrives
to the expression for the Wilson line function ({\it c.f.} Eq.
(\ref{it1c})) (see Fig. \ref{fig7})
\begin{eqnarray}
W(\g,b^2_\perp) &=&   n_c
\Big\{\frac{4}{9}w_c(\g,b^2_\perp)\({\mathbb I}\times {\mathbb
I}\)+
\frac{1}{8}\[\frac{1}{3}w_c(\g,b^2_\perp)+w_s(\g,b^2_\perp)\]\(\l^A\times\l^A\)\Big\}
\  ,
\label{Tc}\\
w_c(\g,b^2_\perp)&=&\int \! d^4 z_0 \[\cos \ \alpha (v_1, z_0)-1\]\[\cos \ \alpha (v_2, z_0)-1\]
\ ,\label{wcSc}\\
w_s(\g,b^2_\perp)&=&-\int \! d^4 z_0 (\hat n^a_1\hat n^a_2)\sin \
\alpha (v_1, z_0) \sin \ \alpha (v_2, z_0) \ ,
\nonumber\end{eqnarray} where the color correlation factor is \be
\hat n^a_1\hat n^a_2 = \frac{(v_1v_2)(z_0,z_0-b_\perp) -
(v_1z_0)(v_2z_0) } {s(v_1,z_{0})s(v_2,z_{0})}\ . \ee The phases
are defined as
\begin{eqnarray}
\alpha (v_1, z_0) &=&   s(v_1,z_{0}) \int_{-\infty}^\infty \!
d\lambda \ \vf\[( z_0+v_1\lambda )^2; \r \]\ ,
\label{it22}\\
\alpha (v_2, z_0) &=&   s(v_2,z_{0}) \int_{-\infty}^\infty \!
d\lambda \ \vf\[(z_0 -v_2\lambda -b_\perp)^2; \r \] \ .
\end{eqnarray}
with $$s^2(v_1,z_{0})=z_0^2-(v_1z_0);\ \ \ \ \
s^2(v_2,z_{0})=(z_0-b_\perp)^2-(v_2z_0).$$ By means of the change
of variables $(z_4\cos\g-z_3\sin\g)\to z_4$, the energy dependence
is trivially factorized \be
w_c(\g,b^2_\perp)\to\frac{1}{\sin{\g}}w_c(\g=\pi/2,b^2_\perp) \ ,\
\ \ \ w_s(\g,b^2_\perp)\to{\cot\g}\ w_s(\g=\pi/2,b^2_\perp)\ .
\label{Gfac}\ee At $\g=\pi/2$ the above definitions are reduced to
\begin{eqnarray}
s^2(v_1,z_{0})\to s_1^2&=&z_3^2+z_\perp^2,\ \
s^2(v_2,z_{0})\to s_2^2=z_4^2+(z-b)_\perp^2, \nonumber\\
\hat n^a_1\hat n^a_2(\g=\pi/2) &=& \frac{z_\perp^2-(zb_\perp) }{s_1s_2}\ ,\nonumber\\
\alpha (v_1, z_0)\to \alpha_1&=&s_1 \int_{-\infty}^\infty \! d\lambda
\ \vf\[z_3^2+z_\perp^2+\lambda^2; \r \]\ ,\nonumber\\
\alpha (v_2, z_0)\to \alpha_2&=&s_2 \int_{-\infty}^\infty \!
d\lambda \ \vf\[z_4^2+(z-b)_\perp^2+\lambda^2; \r \]\ .
\end{eqnarray}

The differential cross section of the quark-quark scattering is
expressed through the amplitude (\ref{Tc}) as \be
\frac{d\sigma_{qq}}{dt}\approx
\frac{1}{9}\frac{1}{s^2}\sum_{kl}\sum_{mn}|T^{km}_{ln}(s,t)|^2 \ .
\label{dsdt}\ee By inserting (\ref{Tc}) into (\ref{dsdt}) and
making analytical continuation to Minkowski space one finds
\begin{eqnarray}
\frac{d\sigma_{qq}(t)}{dt}&=&\frac{2}{9}n_c^2
\[\coth^2\c\ F_s^2(t)+\frac{2}{3}\frac{\coth\c}{\sinh\c} F_c(t)F_s(t)+\frac{11}{3}
\frac{1}{\sinh^2\c}F_c^2(t)\] \ , \label{dsdt1}\end{eqnarray}
where \be F_s(t)=\int d^2\vec b_{\perp} e^{i b_{\perp} q}
w_s(\g=\pi/2,b_\perp^2)\ ,\ \ \ F_c(t)=\int d^2\vec b_{\perp} e^{i
b_{\perp} q} w_c(\g=\pi/2,b_\perp^2)\ . \ee In the asymptotic
limit $(\sinh\c\sim s \ ,\ \coth\c\to1)$ the result (\ref{dsdt1})
coincides with that one found in \cite{Sh} \be
\frac{d\sigma}{dt}\approx \frac{2}{9}n_c^2 F_s^2(t) \ .
\label{dsdtas}\ee In the weak field limit we reproduce the
one-loop single instanton results (\ref{Tmatr}) and (\ref{It}).

Considering the quark-antiquark scattering we have to take into
account that it is possible to treat an antiquark with velocity
$v_2$ as a quark moving backward in time with velocity $-v_2$. As
a result, the scalar product of velocities changes the sign
$(v^q_1v^{\bar q}_2)=-(v^q_1v^q_2)$ and the scattering angles are
related as \be \c_{q q}\to i\pi-\c_{q\bar q}\ . \ee Then one gets
\begin{eqnarray}
\frac{d\sigma_{q\bar q}(t)}{dt}&=&\frac{2}{9}n_c^2
\[\coth^2\c\ F_s^2(t)-\frac{2}{3}\frac{\coth\c}
{\sinh\c}F_c(t)F_s(t)+\frac{11}{3}\frac{1}{\sinh^2\c}F_c^2(t)\] \
. \label{dsdt1qbq}\end{eqnarray} The second terms in (\ref{dsdt1})
and (\ref{dsdt1qbq}) corresponding to the contribution of the
$C-$odd amplitude has been missed in \cite{Sh} because there only
$SU_c(2)$ QCD is considered.

The spin averaged total quark-quark cross section in the
instanton--antiinstanton approximation reads \be
\sigma_{qq}\approx \frac{2}{9}n_c^2\int_0^\infty d\vecc q^2_\perp
\[F_s^2(\vecc q^2_\perp)
+ \frac{4}{3}\frac{m^2}{s}F_c(\vecc q^2_\perp)F_s(\vecc
q^2_\perp)\] \ , \label{Sas}\ee which is constant in the high
energy $s$ limit. It is finite if the constraint instanton
solution is used. In Eq. (\ref{Sas}), the only term corresponding
to the $C=+1$ exchange, Pomeron, survives at asymptotic, while the
$C=-1$ contribution, odderon, (second term in (\ref{Sas})) is
suppressed by the small factor $\sim m^2/s$. The growing part of
the total cross section arises as
\begin{equation}
\Delta\sigma _{qq}\sim (n_{c}\rho _{c}^{4})^2\Delta (t)\ln s \ ,
\label{Sgrow}\end{equation}
considering the inelastic
quark - quark scattering in the instanton-antiinstanton background \cite{Sh}.

As it was discussed in detail in \cc{LN, Sh}, the model of the
Pomeron based on nonperturbative gluon exchanges explains many
properties of the diffractive scattering: the effective
vector-like exchange (\ref{Tmatr}), the additive quark rule and
the main features of the total cross section (\ref{Sas}),
(\ref{Sgrow}).

It is important to note that the original Wilson loop (\ref{1ab})
has essentially Minkowskian light-cone geometry whereas the
instanton calculations of Wilson loop are performed in the
Euclidean QCD. In analogy with the quark form factor the
analytical continuation from Minkowski space to Euclidean one and
{\it v. v.} becomes possible since the dependence of the Wilson
loop on the total energy $s$ and transverse momentum squared $t$
is factorized in (\ref{Tmatr}) and (\ref{Gfac}). At high energy,
the amplitude is $s-$ independent both in the perturbative and
nonperturbative cases. At the same time, the $t-$dependence of the
amplitude is naturally expressed through the nonperturbative
instanton field. Notice that in the original expression for the
Wilson loop (\ref{1ab}), the nonlocal instanton correlator was
integrated over both space-like and time-like separations $x^{2}$
corresponding to the distance between different points on the
contour $C_{\mathrm{qq}}$, whereas the final expressions
(\ref{It}), (\ref{dsdtas}) depend on the space-like variable $t$.
Thus we can proceed the formal calculations in Minkowski space and
then make the Wick rotation $\vecc{k}_{\perp }^{2}\rightarrow
-\vecc{k}_{\perp }^{2}$, or, that is more natural from the point
of view the instanton model, to perform formal manipulations in
the Euclidean space and than make analytical continuation.

\vspace{.3cm}

\section{Conclusions}

Besides the considerable progress in investigation of the role of
nonperturbative QCD vacuum structure (in particular, of the
instanton  phenomena) in low and moderate energy domains of
hadronic physics, nowadays there is a lack of understanding of
their role in high energy processes which are intensively studied
in modern experiments in particle physics. In this work we
presented the results of analysis of the structure of
nonperturbative corrections in such important quantities as the
quark form factor and the cross section of the diffractive
quark-quark scattering at high energy. The quark scattering
process was considered in background of QCD vacuum which is
described within the instanton liquid model. The instanton
contribution to the electromagnetic quark form factor is
calculated in the large momentum transfer regime. We estimated
analytically the weak field approximation for the instanton
contribution and find the all-order result in the asymptotic
regime. The latter was performed considering the Gaussian
simulation of the constraint instanton profile function. We found
that the leading contribution to high energy asymptotic behavior
is provided by the lowest order terms. Although the latter result
could be treated as an argument in favor of validity of the weak
field approximation, the better work has to be done in this
direction since the results for Gaussian profile and the instanton
in the singular gauge may be different in general.

The instanton  effects are more important for phenomenology of the
hadronic processes possessing two different energy scales. (For
more detailed discussions see the works \cite{D00}). One of such
situations---quark-quark diffractive scattering---was considered
in the last Section of the present work. This approach supports
the model of the Pomeron as exchange by nonperturbative gluons
interpolated by the classical instanton field. We have shown that
the $C-$odd (odderon) amplitude is suppressed as $1/s$ compared to
the $C-$even (Pomeron) one. In the case of diffractive scattering,
the total center-of-mass energy $s$ (hard characteristic scale) is
large while the squared momentum transfer $-t$ is small compared
to the latter: $-t \ll s$, but nevertheless larger than any IR
scale. Besides this, the other cases of interest where the
nonperturbative (including instanton induced) effects may be
significant are the saturation in deep-inelastic scattering at
small-x \cc{SCHUT}, and the transverse momentum distribution of
vector bosons in the Drell-Yan process \cc{KRREN}. The latter is
one of the most important objects of the experimental
investigations (in particular, in the context of searches for New
Physics and Higgs bosons---at future LHC and Tevatron experiments
\cc{LHC}), as well as theoretical studies of the predictive power
of pQCD at various energy scales and the role of nonperturbative
physics (see, {\it e.g.}, \cc{DYQIU} and references therein).

\section{Acknowledgements}
The useful discussions on various aspects of this work and
critical comments by B.I. Ermolaev, N.I. Kochelev, E.A. Kuraev, L.
Magnea, S.V. Mikhailov, N.G. Stefanis, and O.V. Teryaev are
thanked. The work is partially supported by RFBR (Grant nos.
04-02-16445, 03-02-17291, 02-02-16194), Russian Federation
President's Grant no. 1450-2003-2, and INTAS (Grant no.
00-00-366).

\section{Appendix}
The evaluation of the path integrals in the weak-field (one-loop)
approximation was performed by using the following expressions:
The partial derivatives in $n$-dimensional space-time are
presented as \be \pd_z^2 = 2 n \pd_{z^2} + 4 z^2 \pd^2_{z^2} \ ,\
\ \ \ \
 \pd_\m \pd_\n = 2 g_{\m\n} \pd_{z^2} + 4 z_\m z_\n
\pd^2_{z^2} \ . \label{A1}\ee The scalar products of the
scattering vectors (\ref{p1p2}) in terms of the scattering angle
$\chi$ read
$$ v^1_\m v^2_\n z_\m z_\n = z^2 \ \cosh \c + \s \t \
\sinh^2 \c \ , $$ \be  z^2 = \(v_1 \s + v_2 \t \)^2 = \s^2 + \t^2
+ 2 \s\t \cosh \c .\label{A2}\ee To extract the universal cusp
factor, we use the integrals:  \be \int_0^{\infty} \!
d\s\int_0^{\infty} d\t \ \exp\[-\a (\s^2 + \t^2 + 2 \s\t \cosh
\c)\] = {1 \over 2 \a}{\c \over \sinh \c} \  \label{Ias1}\ee and
\be \int_0^{\infty} \! d\s\int_0^{\infty} d\t \ \s \t \ \exp\[-\a
(\s^2 + \t^2 + 2 \s\t \cosh \c)\] = - {1 - \c \coth \c \over 4
\a^2 \sinh^2 \c} \ . \label{Ias2}\ee

Therefore, for arbitrary functions $\D_i(z^2)$ we get: \be
\D_i^{\{k\}}(z^2) = (-)^k \int_0^{\infty} \! d\a \a^k \ex^{-\a
z^2} \bar \D_i(\a) \  \ee and gets
$$\int_0^{\infty} \! d\s\int_0^{\infty} d\t \ \D' (z^2) = - {\c \over 2 \sinh \c} \D(0) \ , $$
\be \int_0^{\infty} \! d\s\int_0^{\infty} d\t \ z^2 \D'' (z^2) =
{\c \over 2 \sinh \c} \D(0) \ , \label{A6}\ee
$$ \int_0^{\infty} \! d\s\int_0^{\infty} d\t \ \s\t \D'' (z^2) = - {1 - \c \coth \c \over 4 \sinh^2
\c} \D(0) \ . $$
\vspace{0cm}

In the all-order instanton calculations, the Gaussian integrals
over $z_{0}$ are taken by \be \int d^{2}z_{\perp }\left( z_{\perp
}^{2}\right) ^{n}e^{-\alpha z_{\perp }^{2}}=n!\frac{\pi }{\alpha
^{n+1}} \ , \label{GI1}\ee \be \int_{-\infty }^{\infty
}dyy^{n}e^{-py^{2}- q y}= \sqrt{\frac{\pi}{p}}e^{ \frac{q^2}{4p}}
\sum_{s=0}^{\left[ n/2\right] }  \overline{C}_{s}^{n} \left(
-\frac{q}{2p}\right)^{n-2s} \frac{1}{p^s}, \label{GaussInt} \ee
where
$$ \overline{C}_{s}^{n}=\frac{n!}{s!(n-2s)!2^{2s}} $$
and we use the binomial formula \be \left( z_{3}^{2}+z_{\perp
}^{2}\right) ^{n}=\sum_{k=0}^{n}C_{k}^{n}z_{\perp
}^{2k}z_{3}^{2(n-k)} \ . \ee

\vfill

\eject

\begin{figure*}
\centering \epsfig{file=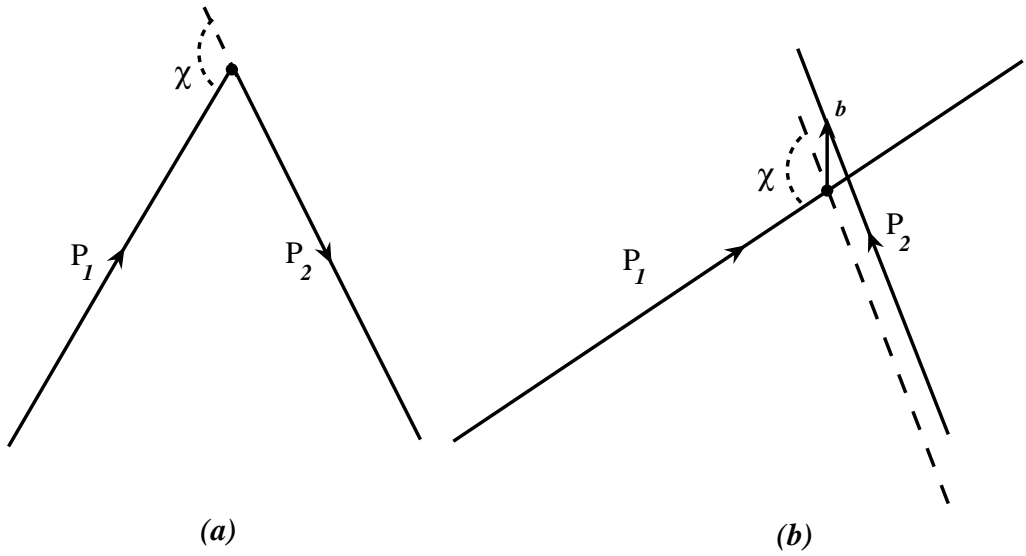,width=0.9\hsize} \caption{The
notations of the quark momenta for the (a) quark form factor, and
(b) quark-quark scattering.} \label{nots}
\end{figure*}

\begin{figure*}
\centering \epsfig{file=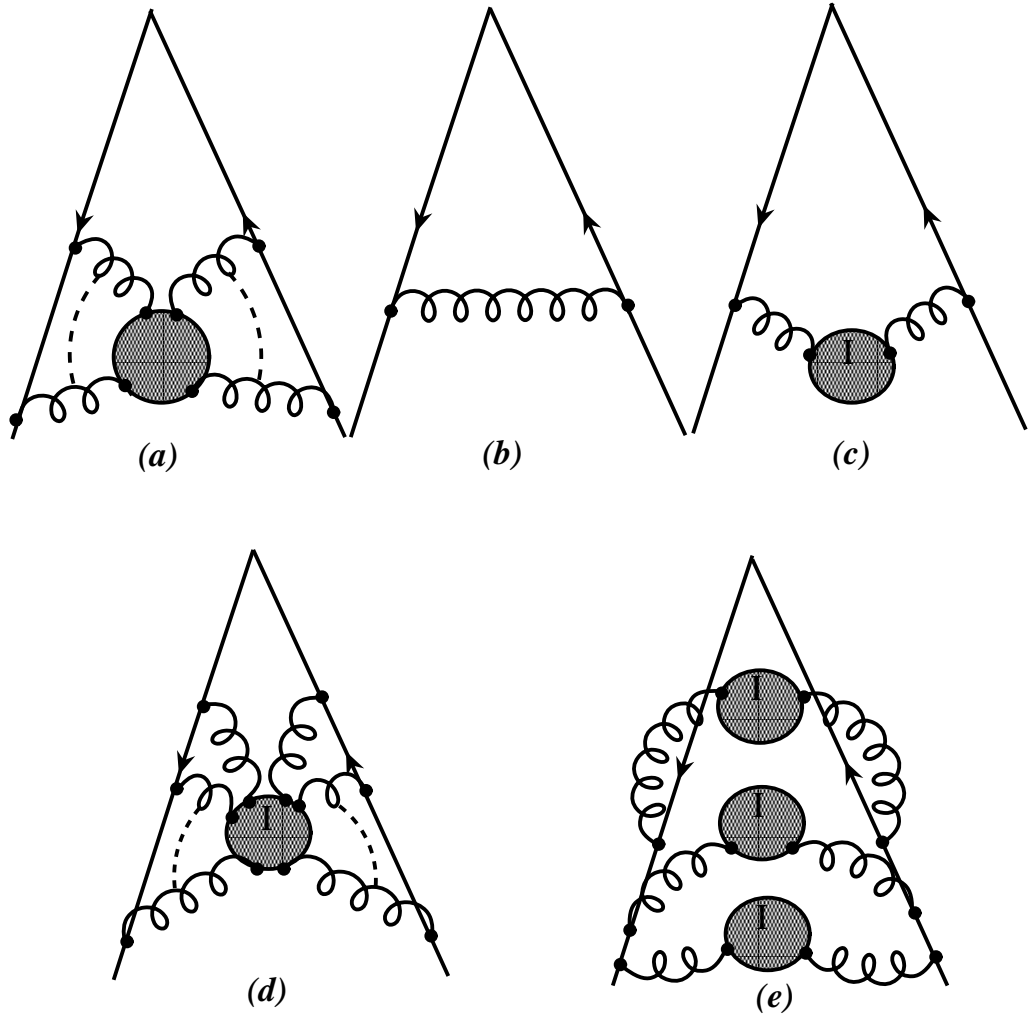,width=0.9\hsize} \caption{The
total cusp-dependent part of the Wilson loop integral for the
quark form factor (a); the leading order contributions of the
perturbative (b) and nonperturbative (single-instanton) (c)
fields; (d) the all-order single instanton result; (e) the
exponentiation of the single instanton result.} \label{fig0}
\end{figure*}

\begin{figure*}
\centering \epsfig{file=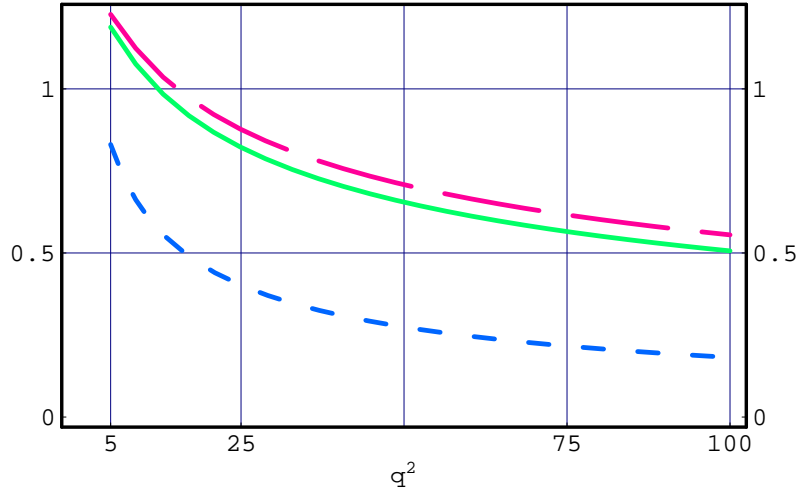,width=0.7\hsize} \caption{The
asymptotic behaviour of the quark form factor is shown as the
function of the dimensionless variable $q^2 = Q^2/\L^2$, up to
terms $O\(\ln \ln q^2\)$. The long-dash presents the contribution
of one loop perturbative terms; the solid line represents the
total form factor including the instanton induced part, Eq.
(8.21). For comparison, the leading $(\sim  \ln q^2 \ln \ln q^2)$
perturbative contribution is shown separately---the short-dash
line.} \label{fig1}
\end{figure*}

\begin{figure*}
\centering \epsfig{file=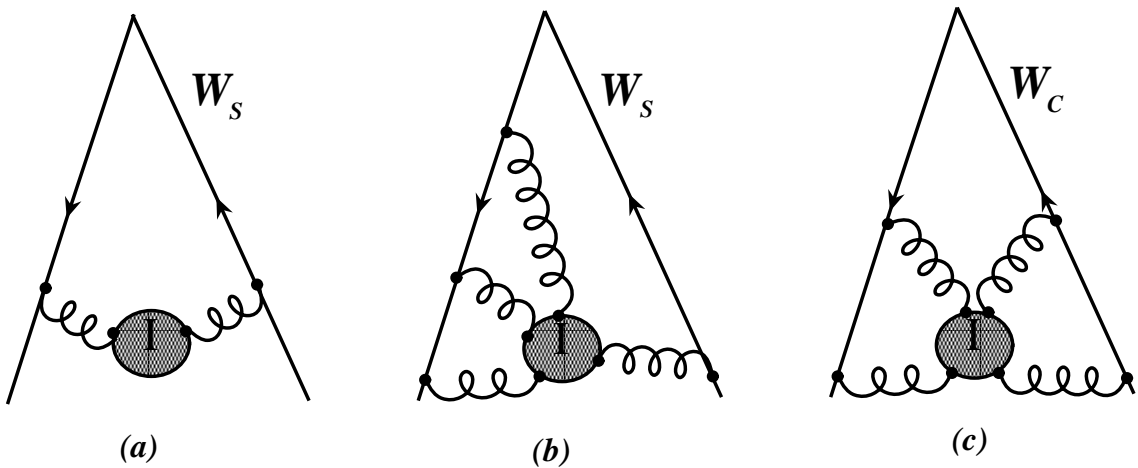,width=1.0\hsize}
\caption{Schematic representation of the lowest order instanton
contributions.} \label{fig3}
\end{figure*}

\begin{figure}
\includegraphics[width=10cm,height=7cm]{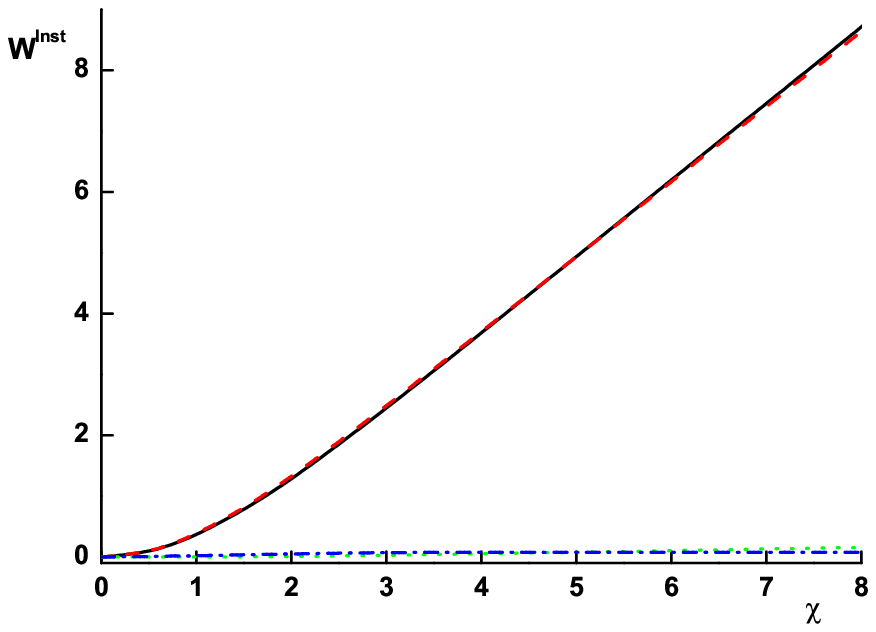}
\caption{Lowest orders instanton contributions to the Wilson
integral with spin factors. The leading term $\left\langle \alpha
_{1}^1\alpha _{2}^1\right\rangle(\gamma)$ is the dashed line,
next-to-leading terms $\left\langle \alpha _{1}^1\alpha
_{2}^3\right\rangle(\gamma)$ and $\left\langle \alpha _{1}^2\alpha
_{2}^2\right\rangle(\gamma)$ are the dotted and dash-dotted line,
correspondingly. The sum of these contributions is the solid
line.} \label{f2}
\end{figure}

\begin{figure}[]
\includegraphics[width=10cm,height=7cm]{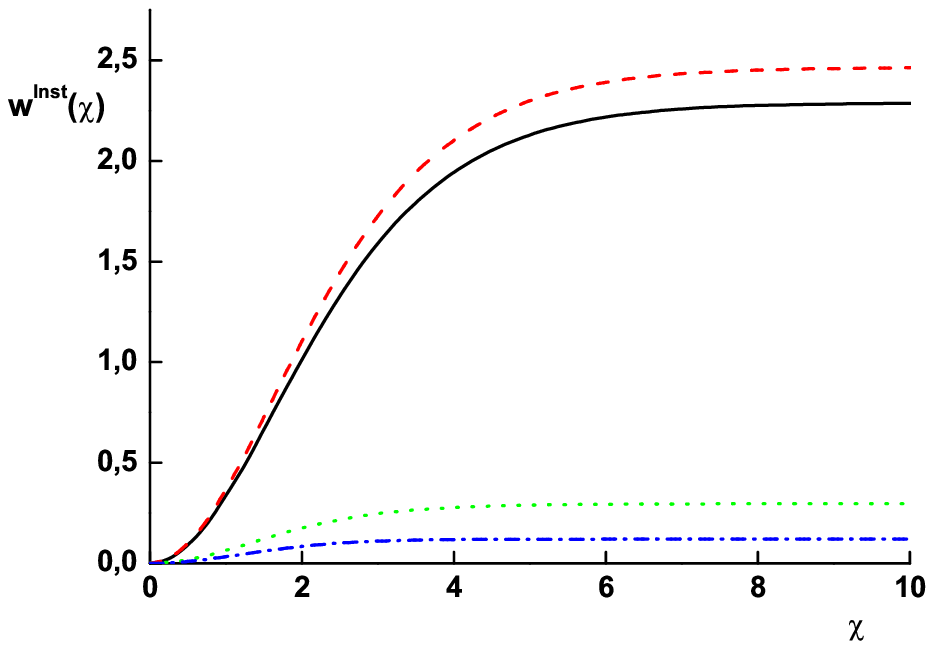}
\caption{Lowest orders instanton contributions to the Wilson
integral without spin factors. The leading term $\left\langle
\alpha _{1}^1\alpha _{2}^1\right\rangle(\gamma)$ is dashed line,
next to leading terms $\left\langle \alpha _{1}^1\alpha
_{2}^3\right\rangle(\gamma)$ and $\left\langle \alpha _{1}^2\alpha
_{2}^2\right\rangle(\gamma)$ are dotted and dash-dotted line,
correspondingly. The sum of these contributions is the solid
line.} \label{f3}
\end{figure}

\eject

\begin{figure*}
\centering \epsfig{file=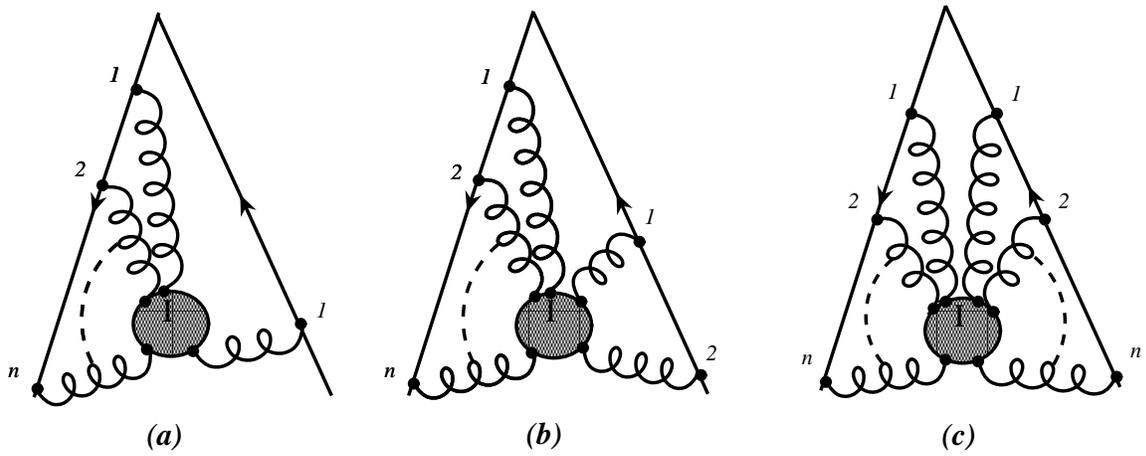,width=1.\hsize}
\caption{Schematic representation of partial summation of the
instanton contributions.} \label{fig6}
\end{figure*}
\begin{figure*}
\centering \epsfig{file=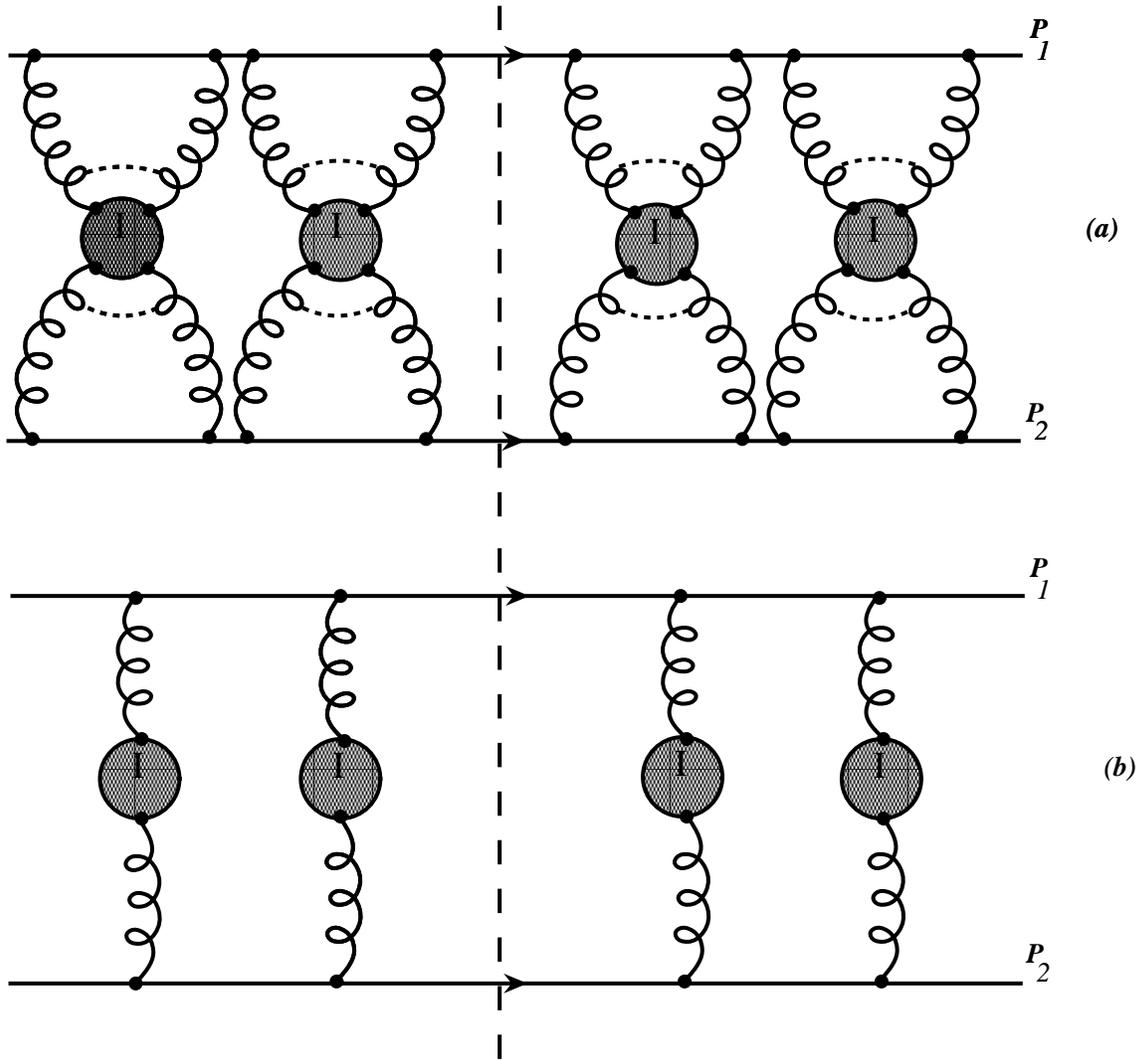,width=1.\hsize}
\caption{Schematic representation of partial summation of the
instanton contributions to the quark-quark scattering.}
\label{fig7}
\end{figure*}
\eject
\begin{figure*}
\centering \epsfig{file=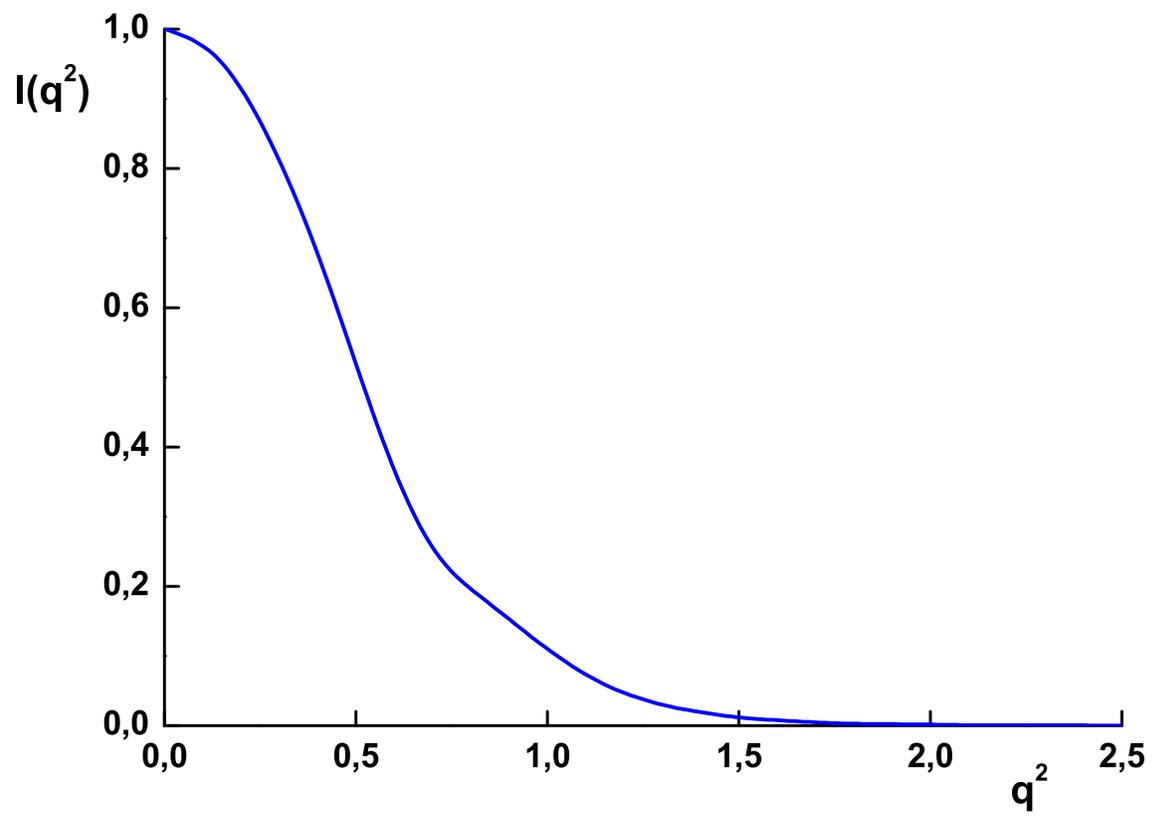,width=1.\hsize}
\caption{Normalized quark-quark scattering amplitude, Eq.
(10.13).} \label{fig8}
\end{figure*}

\end{document}